\documentclass[aps, prc, reprint, twocolumn, superscriptaddress]{revtex4-1}

\usepackage{graphicx}
\usepackage{dcolumn}
\usepackage{bm}
\usepackage{hyperref}
\usepackage[mathlines]{lineno}
\usepackage{color}
\usepackage{longtable}

\begin{document}

\title{Cross section measurements of $^{155,157}$Gd(n,$\gamma$) induced by thermal and epithermal neutrons}
\author{M.~Mastromarco} \affiliation{Istituto Nazionale di Fisica Nucleare, Sezione di Bari, Italy} %
\author{A.~Manna} \affiliation{Istituto Nazionale di Fisica Nucleare, Sezione di Bologna, Italy} \affiliation{Dipartimento di Fisica e Astronomia, Universit\`{a} di Bologna, Italy} %
\author{O.~Aberle} \affiliation{European Organization for Nuclear Research (CERN), Switzerland} %
\author{S.~Amaducci} \affiliation{Istituto Nazionale di Fisica Nucleare, Sezione di Bologna, Italy} \affiliation{Dipartimento di Fisica e Astronomia, Universit\`{a} di Bologna, Italy} %
\author{J.~Andrzejewski} \affiliation{University of Lodz, Poland} %
\author{L.~Audouin} \affiliation{Institut de Physique Nucl\'{e}aire, CNRS-IN2P3, Univ. Paris-Sud, Universit\'{e} Paris-Saclay, F-91406 Orsay Cedex, France} %
\author{M.~Bacak} \affiliation{Technische Universit\"{a}t Wien, Austria} \affiliation{European Organization for Nuclear Research (CERN), Switzerland} \affiliation{CEA Irfu, Universit\'{e} Paris-Saclay, F-91191 Gif-sur-Yvette, France} %
\author{J.~Balibrea} \affiliation{Centro de Investigaciones Energ\'{e}ticas Medioambientales y Tecnol\'{o}gicas (CIEMAT), Spain} %
\author{M.~Barbagallo} \affiliation{Istituto Nazionale di Fisica Nucleare, Sezione di Bari, Italy} %
\author{F.~Be\v{c}v\'{a}\v{r}} \affiliation{Charles University, Prague, Czech Republic} %
% ...
%
\collaboration{The n\_TOF Collaboration (www.cern.ch/ntof)} \noaffiliation 
\date{\today}
\begin{abstract}
Neutron capture measurements on $^{155}$Gd and $^{157}$Gd were performed using the time-of-flight technique at the n\_TOF facility at CERN. Four samples in form of self-sustaining metallic discs isotopically enriched in $^{155}$Gd and $^{157}$Gd were used. The measurements were carried out at the experimental area (EAR1) at 185 m from the neutron source, with an array of 4 C$_6$D$_6$ liquid scintillation detectors. 

The capture cross sections of $^{155}$Gd and $^{157}$Gd at neutron kinetic energy of 0.0253 eV have been estimated to be 62.2(2.2) kb and 239.8(9.3) kb, respectively, thus up to 6\% different relative to the ones reported in the nuclear data libraries.  
A resonance shape analysis has been performed in the resolved resonance region up to 180 eV and 300 eV, respectively, in average resonance parameters have been found in good agreement with evaluations. Above these energies the observed resonance-like structures in the cross section have been tentatively characterised in terms of resonance energy and area up to 1 keV.
\end{abstract}
\maketitle
\section{Introduction\label{Introduction}}
The natural element with the highest cross section for thermal neutrons is Gadolinium. Among its 7 stable isotopes, $^{157}$Gd and, to a smaller extent, $^{155}$Gd are responsible for this feature, mainly attributable to the existence of a neutron resonance near thermal energy (i.~e. neutron kinetic energy $E_n=0.0253$ eV, corresponding to a speed of 2200 m/s). Accurate values of the neutron capture cross section of gadolinium isotopes are required in many fields of science: for the understanding of the nucleosynthesis of elements beyond iron in stars (via the $\small s$ and $\small r$ processes)~\cite{sproc}; for the neutron capture therapy of cancer~\cite{DeAgo}; for the development of neutrino detectors~\cite{BEACOM} and for nuclear technologies. In the last case, it has a relevant role in the neutron balance and the safety features of Light Water Reactors and Canada deuterium uranium (CANDU) reactor types, since gadolinium is used as the burnable poison in the fuel pin or  moderator in the reactor core~\cite{Rocchi}.

The $^{155}$Gd(n,$\gamma$) and $^{157}$Gd(n,$\gamma$) cross sections are available in nuclear data libraries such as ENDF/B-VIII.0~\cite{ENDFB8}, 
JEFF-3.3~\cite{JEFF33}, and JENDL-4.0~\cite{JENDL40}. In particular, ENDF/B-VIII.0 is taken over from the previous ENDF/B evaluation, which is based on the resonance parameters compiled by Ref.~\cite{MUGH}. The resonance parameters for $^{157}$Gd are consistent with those from the experiment by M{\o}ller and collaborators~\cite{MOLLER} and not with those from Leinweber and collaborators~\cite{LEINW}. JEFF-3.3 is taken from JEFF-2, which adopts: JENDL-2 for $^{155}$Gd; and JENDL-1 for $^{157}$Gd. The latter is based on the resonance parameters from the BNL-325 report (the previous edition of Ref.~\cite{MUGH}). In JENDL-4.0, the $^{155,157}$Gd evaluations were revised by considering resonance parameter from Ref.~\cite{LEINW}. In the case of $^{157}$Gd, a background capture cross section was added below 0.1 eV in order to reproduce the thermal cross section of the previous evaluation and a number of benchmarks reported in the International Criticality Safety Benchmark Evaluation Project~\cite{ICSBEP} (more details in~\cite{JENDL40}).  In summary, evaluations agree on the adoption of the $^{157}$Gd(n,$\gamma$) thermal cross section consistent with the experiment by Ref.~\cite{MOLLER}, although that value is about 12\% higher than what was measured by Leinweber and collaborators~\cite{LEINW} in a capture and transmission experiment.  
A summary of the values of the thermal cross sections retrieved from the experimental nuclear reaction database EXFOR  is reported in historical progression in Tab.~\ref{tab:csth}. The measurements of Pattenden~\cite{Pattenden}, Tattersall~\cite{Tattersall} and Choi~\cite{CHOI} are not listed because they are not direct measurements and depends on model calculations.
\begin{table}[htbp]
\caption{$^{155}$Gd and $^{157}$Gd thermal cross sections (in kb) as reported in literature,  compilation~\cite{MUGH} and evaluations.\label{tab:csth}}
\begin{ruledtabular}
\begin{tabular}{lccc}
Reference & Year & \multicolumn{2}{c}{Thermal cross section} \\
                  &         & n $+^{155}$Gd & n $+^{157}$Gd\\
\hline
M{\o}ller~\cite{MOLLER} & 1960 & $58.9(5){}^a$&$254(2){}^a$\\
Ohno~\cite{OHNO} & 1968 & $61.9(6){}^a$& $248(4){}^a$\\
Leinweber~\cite{LEINW} & 2006 & 60.2$^b$&226$^b$\\
Mughabghab~\cite{MUGH} & 2009&$60.9(0.5)$ &$254.0(0.8)$\\
JENDL-4.0 & 2016 & 60.735&253.25 \\
JEFF-3.3 & 2017 & 60.89&254.5\\
ENDF/B-VIII.0 & 2018 & 60.89&253.32\\
 \end{tabular}
 \end{ruledtabular}
 $^a$ Total cross section.\\
 $^b$ The uncertainty is not explicitly quoted in Ref.~\cite{LEINW}.\\
  \end{table}

 At higher energies, up to about 300 eV and especially beyond 180 eV (where the 
evaluations of $^{155}$Gd terminate) isotopic resonance assignement in the literature are not consistent and thus evaluations reflect these doubtful assignments. 
Moreover, the resonance parameters deduced from the measurement of Leinweber 
and collaborators~\cite{LEINW} are significantly different from the 
ENDF/B-VIII.0 and JEFF-3.3 evaluations. However, this large deviation is not completely confirmed by a capture measurement on $^{155}$Gd~\cite{BARA}.
 In summary, a substantial improvement of the parametrization
 of $^{155}$Gd(n,$\gamma$) and $^{157}$Gd(n,$\gamma$) cross sections seems necessary.  

All these inconsistencies prompted an initiative to perform a new measurement of the capture cross section for the
odd gadolinium isotopes in the resolved resonance region at the CERN neutron time-of-flight facility, n\_TOF. 
In this work, we report the results of these measurements on $^{155}$Gd and $^{157}$Gd. In Sec.~\ref{Sec_2} the experimental conditions are discussed, while Sec.~\ref{Sec_3} describes the data reduction procedure together with the estimation of uncertainties. Section~\ref{Sec_4} summarizes the neutron resonance analysis: results and discussions for $^{155}$Gd(n,$\gamma$) and $^{157}$Gd(n,$\gamma$) are reported in Sec.~\ref{sec155} and ~\ref{sec157}, respectively. A resonance analysis above the resolved resonance region is reported in Appendix~\ref{App:newres}.    
\section{Capture experiment\label{Sec_2}}
The $^{155}$Gd(n,$\gamma$) and $^{157}$Gd(n,$\gamma$) capture measurements were performed in 2016 at the 185-m measurement station of the neutron time-of-flight facility n\_TOF~\cite{nTOF} at CERN, using an array of 4 C$_6$D$_6$ detectors. The enriched gadolinium samples were in the form of self-supporting metal discs.  Since the
capture cross section for both isotopes drops by several orders of magnitude for neutron energies higher than 1 eV, two samples (thin and thick) for each isotope were used to properly perform the measurement in the whole energy range of interest, i.~e. up to 1 keV.
\subsection{The n\_TOF spectrometer}
The  n\_TOF facility features two beam lines with a white neutron spectrum produced by spallation induced by 20 GeV/c protons impinging on a massive lead target, 40 cm in length and 60 cm in diameter. This experiment was performed at the experimental area EAR1, at the nominal distance of 185 m from the neutron-producing target, because of the better resolving power of the spectrometer.   
Pulses of $7\times10^{12}$ protons impinge on the spallation target, producing some $2\times10^{15}$ neutrons. Those are collimated towards EAR1 to a 0.9-cm-radius beam at 178 m from the neutron source. Consequently the reduction of the neutron intensity attributable to the solid angle subtended by the collimator is of the order of $10^{-9}$. The initially fast neutron spectrum is moderated by a first layer of 1 cm of demineralized water plus a second layer of 4 cm of borated water (H$_2$O + 1.28\%H$_3$BO$_3$, fraction in mass). With this setup the energy spectrum ranges from thermal energies up to the GeV region, a more detailed description of the neutron flux is reported in Sec.~\ref{Subsec_FLUX}. Neutron pulses are produced with a slightly varying frequency of about 1 Hz, thus preventing the overlap of slow neutrons of a bunch with the next bunch. 
The relative energy resolution $\Delta E_n/E_n$, with $\Delta E$ being the full width at half maximum, is of the order of $3.2\times10^{-4}$ at 1 eV and $5.4\times10^{-4}$ at 1 keV (more details in Ref.~\cite{nTOF} and~\cite{G4}). Therefore the resolution of the n\_TOF spectrometer results to be smaller than the total width of neutron resonances in Gd up to about $E_n=250$ eV. On the other hand, for the Doppler broadening (related to the thermal motion of the atoms in the sample) FWHM $\approx 6, 150, 300$ and 600 meV at $E_n=0.02, 12.5, 50$ and 200 eV, respectively, and thus starts to dominate the width of the observed resonance profile above 200 eV.       
\subsection{Capture detectors and instrumentation} 
The measurements were performed using an array of four deuterated benzene (C$_6$D$_6$) liquid scintillation detectors (volume of about 1 liter).
These detectors, widely recognized~\cite{NDS} to be particularly suited for (n,$\gamma$) measurements, were further optimized~\cite{PFM}. so as to have a very low sensitivity to background signals induced by sample-scattered neutrons (i.~e. the amount of material constituting the detectors was minimized, and the materials used have a very low neutron capture cross-section). They were placed face to face at 90$^o$ with respect to the beam and about 10 cm away from the sample.

The total energy detection principle was used by combining the detection system described above with the so called Pulse Height Weighting Technique (PHWT). More details are presented in Sec.~\ref{WF}. 

The data acquisition system consisted of 14 bit flash-ADC channels of TELEDYNE SP-Devices. These devices are equipped with an on-board memory of 512 MB per channel and can record digitized signals for 100 ms corresponding to all neutron energies down to 18 meV for EAR1. For the Gd measurement campaign, 4 channels for the C$_6$D$_6$ signals at a sampling rate of 1 GSample/s and 4 channels for the flux detectors with a sampling rate of 62.5 MSample/s were used.
\subsection{Neutron flux\label{Subsec_FLUX}}
The procedure for the characterization and the definition of the so-called evaluated neutron flux is described in detail in Ref.~\cite{BARB}. It results from a combination of dedicated measurements performed with different detectors, based on neutron cross-section standards~\cite{CARL}. One of these detectors, the silicon monitor (SiMON)~\cite{SiMon} was used during the measurement campaign to keep the neutron flux under systematic control. SiMON is based on the $^6$Li(n, t)$\alpha$ standard and consists of a 600 $\mu$g/cm$^2$ LiF foil in the beam, viewed by 4 silicon detectors (5 cm$\times5$ cm$\times 300$~$\mu$m) out of the beam. Such a configuration makes the SiMON apparatus almost transparent to neutrons. For instance the correction for the presence the LiF foil absorbs less than 1.4\% of thermal neutrons passing through it, moreover the reduction of the incoming neutron beam decreases with increasing neutron energy and becomes negligible (neutron transmission $>99.5\%$) for neutron energies higher than 0.2 eV. Nevertheless the correction was applied in the whole energy region of interest.
 
\begin{figure}
\includegraphics[width=0.48\textwidth]{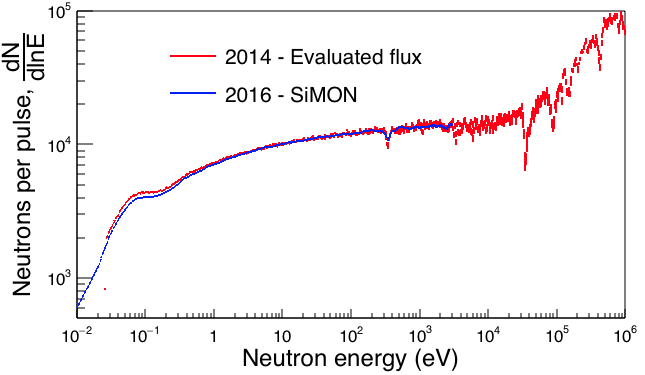}
\caption{(Color online) Neutron flux measured during the gadolinium campaign in the
range 10 meV - 3 keV, compared to the 2014 evaluated flux.\label{FIG_FLUX}}
\end{figure}

Fig.~\ref{FIG_FLUX} shows the energy distribution of the neutron flux at the sample position, for the nominal proton bunch of $7\times10^{12}$ protons, from the 2014 flux evaluation campaign and from the gadolinium campaign. The latter flux was extracted using SiMON and is shown up to E$_n=3$ keV, a region where no sizable correction for non-isotropic emission of the reaction products is required. Above 1 eV the two curves agree within uncertainties, since the shape of the neutron flux is determined by the collimation system (which was not changed). However, in the energy region below 1 eV, a systematic effect as a function of the energy is clearly visible with the deviation reaching 9\% near thermal. This behavior is consistent with a 7\% increase of the concentration of boric acid in the moderator circuit with respect to 2014. 

The 2016 neutron flux, with 100 bins per decade, has been determined within 1\% uncertainty between thermal and
200 eV. The uncorrelated uncertainties, attributable to counting statistics, start to play a major role at higher energies. Therefore, in order to avoid statistical fluctuations due to reduced statistics, the evaluated flux was used above 200 eV.  

As a final remark, the pick-up detection system based on the wall current  monitor of the CERN Proton Synchrotron was used for monitoring the proton current, which is proportional to the proton pulse intensity, and thus, to the neutron beam intensity.
\subsection{Samples\label{samples}}
The gadolinium samples were acquired from the National Isotope Development Center (NIDC) of the Oak Ridge National Laboratory (USA). Since the $^{155,157}$Gd(n,$\gamma$) cross section changes by several orders of magnitude depending on the neutron energy, the measurement in the whole energy range cannot be performed with a single sample. In particular, to avoid saturation of the capture yield attributable to self-shielding, very thin samples (3.2 and 1.6 mg/cm$^2$) were used to measure the cross sections near thermal-neutron energy. For the characterization of resonant structures above 1 eV, 10- and 40-time thicker samples, for $^{155}$Gd and $^{157}$Gd, respectively, were used in order to obtain a good signal-to-background ratio in the resonance region. Hereafter, we will refer to them as to thin and thick gadolinium samples. 
Moreover, the samples were highly isotopically enriched with a cross contamination of the two isotopes of less than 1.14\%. In this way, possible background effects introduced by capture events in the contaminants were minimized. Table~\ref{table:sample} summarizes the characteristics of the samples provided by NIDC, together with the declared uncertainties estimated from an isotopic analysis performed by the provider. The uncertainty on the areal density accounts for both uncertainties related to the weight and the area.

In addition to the 4 gadolinium samples, a $^{197}$Au ($6.28\times10^{-4}$ at/b) and a lead sample ($6.71\times10^{-3}$ at/b) were used for normalization purposes and for the study of the background. 
All the samples were circular in shape with a radius of 1 cm, in the form of self-sustaining metallic discs, thus avoiding background introduced by a sample container.
\begin{table*}[htbp]
\caption{Features of the Gd samples with uncertainties declared by the provider.\label{table:sample}}
\begin{ruledtabular}
\begin{tabular}{cccccc}
Isotope &  abundance &\% contamination &\% main& Weight  & Areal Density \\
 &  \% &of $^{155\mbox{ or }157}$Gd &contaminant &mg &  atoms/barn $\times10^{-8}$\\
\hline
$^{155}$Gd  & $91.74\pm0.18$ &$1.14\pm0.01$  &$5.12\pm0.18$ $^{156}$Gd&100.6 $\pm$ 0.1 & $12438\pm15$\\
$^{155}$Gd  & $91.74\pm0.18$ &$1.14\pm0.01$&$5.12\pm0.18$ $^{156}$Gd&10.0 $\pm$ 0.1 & $1236\pm12$\\
$^{157}$Gd  & $88.32\pm0.01$ & $0.29\pm0.01$ &$9.10\pm0.01$ $^{158}$Gd&191.6 $\pm$ 0.1 & $23390\pm20$ \\
$^{157}$Gd  & $88.32\pm0.01$ & $0.29\pm0.01$ &$9.10\pm0.01$ $^{158}$Gd&4.7 $\pm$ 0.1 & $574\pm12$\\
 \end{tabular}
 \end{ruledtabular}
 \end{table*}

To prevent oxidation of Gd samples, they were shipped in an airtight under-pressurized container. At the beginning of the experimental campaign they were extracted, weighted and sandwiched between two Mylar foils (thickness $ \approx 6 \mu$m) with a small amount of glue. The empty-sample was prepared as a replica of the Gd samples excluding the Gd metal disc. Particular care was given to its production, similar to the gadolinium sample holder ( i.~e. an Al ring with 2 foils of mylar and a layer of glue).

\section{Data analysis \label{Sec_3}}
The fraction of the neutron beam producing a neutron-capture reaction in the sample, namely the capture yield~\cite{NDS} $Y(E_n)$, was obtained from the weighted C$_6$D$_6$ counting rate $C_w$ measured with a Gd sample in the beam:
\begin{equation}\label{eq:yield}
Y(E_n)=\frac{N}{S_n+E_n\frac{A}{A+1}}\frac{C_w(E_n)-B_w(E_n)}{\varphi_n(E_n)f_{BIF}(E_n)},
\end{equation}
where $N$ is a normalization factor independent of neutron energy, $S_n$ is the neutron separation energy of the compound nucleus, $A$ is the mass number of the target nucleus, $B_w$ is the weighted background, $\varphi_n$ is the neutron fluence and $f_{BIF}$ is a correction factor taking into account the variation of the neutron-beam interception as a function of the neutron energy. In the following sections a detailed explanation of the data reduction procedure is addressed, with particular care to the correction factors (Secs.~\ref{WF}, \ref{bkg} and \ref{BIF}) which determine the final uncertainty. The study of the stability of the detectors, their calibration and the time-of-flight to neutron energy conversion are also discussed in sections~\ref{calibration} and \ref{T2E}.
\subsection{Detector resolution and calibration\label{calibration}}
\begin{figure}
\includegraphics[width=0.48\textwidth]{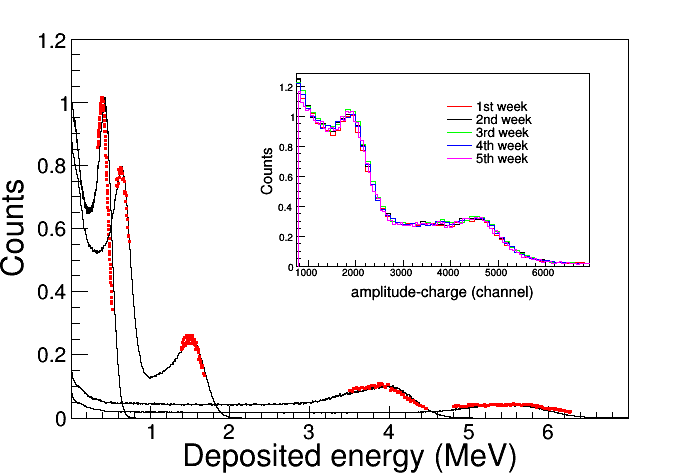}
\caption{Calibrated amplitude spectra for $^{137}$Cs, $^{88}$Y, Am-Be and Cm-C standard $\gamma$-ray sources. The black lines correspond to the simulated spectra convoluted with the energy resolution, the red dots are the experimental data. The inset shows the $^{88}$Y spectra measured during the gadolinium experimental campaign. \label{fig:Exp_Sim2}}
\end{figure}

With the aim of obtaining high accuracy cross section data, the experimental setup has been carefully characterized, with particular care over the stability and the performance of the detectors. The stability of the detector response, mostly related to the gain of the photomultipliers, has been regularly verified by measurements with standard $\gamma$-ray sources, namely  the $^{137}$Cs (E$_\gamma=0.662$ MeV), $^{88}$Y (E$_\gamma=0.898$ MeV and E$_\gamma=1.836$ MeV), and the composite Am-Be (E$_\gamma=4.44$ MeV) and Cm-C (E$_\gamma=6.13$ MeV) sources (see~\ref{fig:Exp_Sim2}).
The energy spectra of $\gamma$-ray sources were recorded more than once per week and did not reveal a gain shift higher than 0.7\%, see the inset of Fig.~\ref{fig:Exp_Sim2} obtained with the Yttrium source.

A reliable calibration of the detectors is an important task in the data analysis, because of the modification of the detector response by means of the PHWT. The calculation of the weighting factors, indeed, depends on the discrimination level applied to the deposited energy spectra. For this reason, particular care was taken to determine the experimental resolution as a function of $\gamma$-ray energy and energy calibration of the detectors. The iterative procedure followed to achieve this objective consisted of (i) extraction of energy resolution from experimental spectra acquired with the standard $\gamma$-ray sources described above; (ii) broadening of the simulated spectra with the energy resolution so as to reproduce the measurements, as shown in Figure ~\ref{fig:Exp_Sim2}; (iii) fine calibration of the detectors determined by the best matching between simulated and measured spectra.
\subsection{Determination of Weighting functions \label{WF}}
The total-energy detection principle was applied for the evaluation of the neutron capture cross-section in the resonance region. The use of low efficiency detection system, such as the present setup based on C$_6$D$_6$ detectors, is at the heart of this approach, along with the adoption of the pulse height weighting technique (see Ref.~\cite{NDS} and ~\cite{BORE} and references therein). The latter ensures proportionality of the $\gamma$-ray detection efficiency to the corresponding  $\gamma$-ray energy. The proportionality is achieved by introducing a mathematical procedure based on a weighting function $WF(E_d)$, $E_d$ being the energy deposited by a $\gamma$ ray. $WF(E_d)$ is defined so that the detection efficiency for the weighted response function is proportional to the energy of the detected $\gamma$ ray. The description of the detection system response was determined using a Monte Carlo simulation of the apparatus, by {\sc geant4} simulation~\cite{GEANT} of the complete experimental assembly. The response of the detection system has been studied as a function of the $\gamma$-ray energy. In the MC simulation, the $\gamma$ rays were emitted from the sample according to the Gaussian $xy$ distribution of the neutron beam profile, uniformly in $z$ direction with $z$ axis being the direction of the neutron beam. Then the final detector response was obtained by convoluting the simulated response with a Gaussian function that represents the detector resolution. 
Since the $\gamma$-ray transport in the sample can play a relevant role for the thick samples and, generally, for high values of the product $n \sigma_{tot}$, where $n$ is the areal density in atoms/barn and $\sigma_{tot}$ is the total cross section. The case of a spatial distribution with exponential shape along $z$ for the emitted $\gamma$ ray in the sample was evaluated and is discussed in Sec.\ref{sec:QA}.

The weighting function $WF(E_d)$ was parameterized with a polynomial function by minimizing the difference between the weighted response and the corresponding $\gamma$-ray energy for a number of energies in the range of interest. 
The discrimination level was fixed to 150 keV and the upper threshold to 10.0 MeV, corresponding to the Compton edge of $\gamma$-ray energies of 285 keV and 10.3 MeV, respectively. The upper threshold exceeds the neutron separation energies of $^{156}$Gd ($S_n=8.54$ MeV), $^{158}$Gd ($S_n=7.94$ MeV) and $^{198}$Au ($S_n=6.51$ MeV) to take into account the resolution broadening of the scintillation detectors. The impact of these analysis conditions on the uncertainty related to the PHWT is discussed in Sec.\ref{sec:QA}.

The loss of cascade $\gamma$-rays attributable to the electron conversion process should be considered for a careful estimation of the uncertainty related to the PHWT.  The Monte-Carlo {\sc dicebox} algorithm~\cite{Becvar97} was used to produce artificial capture cascades and estimate this effect. In {\sc dicebox}, the complete decay scheme is taken from existing experimental data below certain critical energy, $E_{\rm crit}$. The data from Refs.~\cite{NDS156},~\cite{NDS158} and \cite{NDSAu} for $^{156,158}$Gd and $^{198}$Au, respectively, were used and the $E_{\rm crit}$ was adjusted for each isotope to ensure the completeness of the decay scheme. Above $E_{\rm crit}$ the statistical model, in terms of level density (LD) and a set of photon strength functions (PSFs) for different transition types, was adopted to generate individual levels and their decay properties. The LD and PSFs models and their parameters were taken from Refs.~\cite{Chyzh2011},~\cite{Baramsai2013} and \cite{PSFAu} for $^{156,158}$Gd and $^{198}$Au, respectively. {\sc dicebox} computes the contribution of internal electron conversion using parameters from the BrIcc database~\cite{ICC} for all transitions above $E_{\rm crit}$ and for those transitions below $E_{\rm crit}$ where the experimental information about internal electron conversion is lacking.   

\begin{figure}\includegraphics[width=0.48\textwidth]{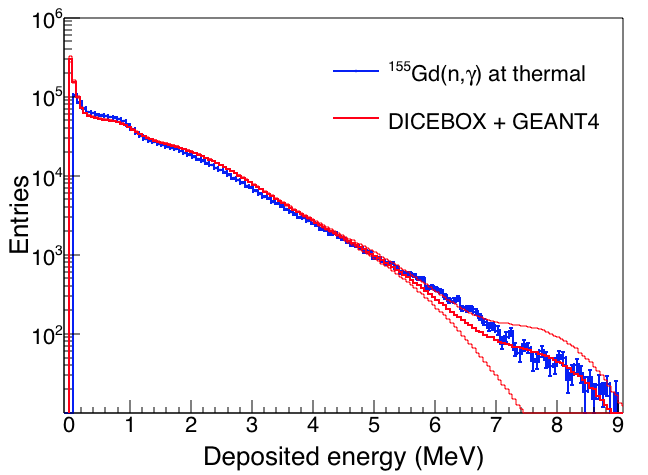}\caption{(Color online) C$_6$D$_6$ energy spectrum for $^{155}$Gd(n,$\gamma$) at neutron energy near thermal compared to the simulated response. \label{FIG-Eg}}\end{figure}

The generated cascades were used as input to the  {\sc geant4} simulation described above. The comparison between experimental and simulated response of the detector is shown in Fig.~\ref{FIG-Eg} for the case of $^{155}$Gd(n,$\gamma$) measurement near neutron-thermal energy.  The red curves (central value and standard deviation) represents the uncertainty in the calculation, which related to the fluctuations in the modeling of the cascades.  The results of this comparison is descibed in Sec.\ref{sec:QA}.

\subsection{Background subtraction\label{bkg}}
\begin{figure}
\includegraphics[width=0.48\textwidth]{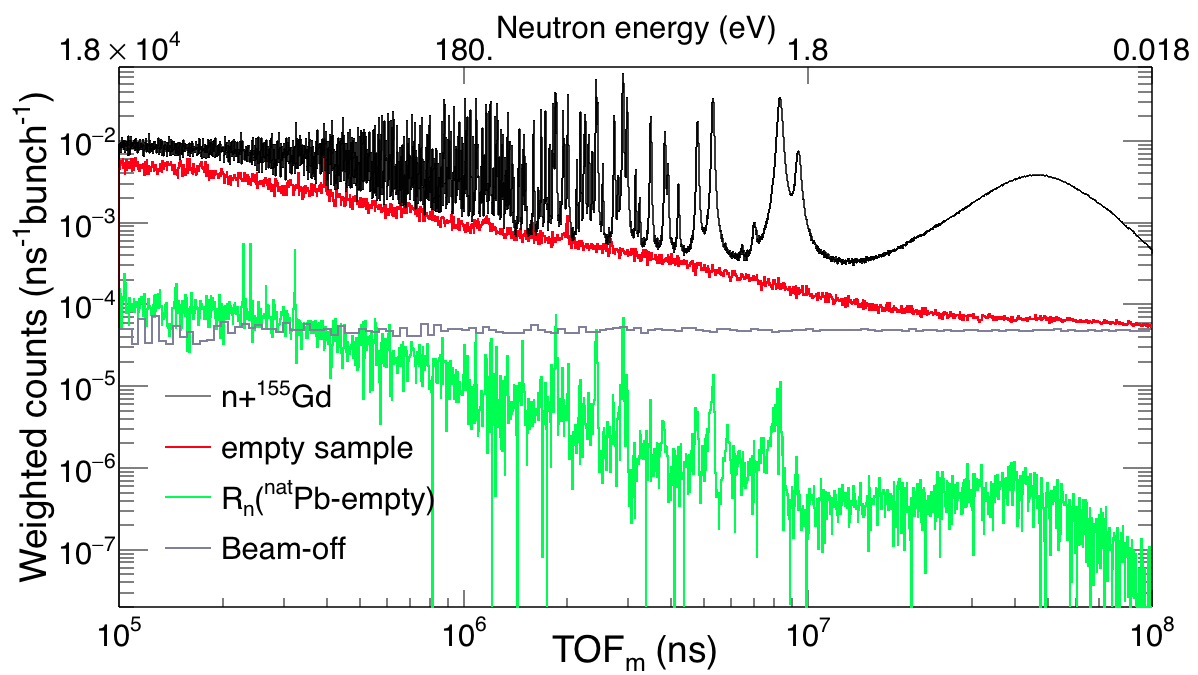}
\caption{(Color online) Weighted C$_6$D$_6$ time-of-flight spectrum of the thick $^{155}$Gd sample, together with background measurements. \label{fig:bkg}}
\end{figure}

The study of the background in the capture data ($B_w$ in Eq.~\ref{eq:yield}) is based on dedicated measurements aimed at evaluating the various components attributable to: (i) neutron beam interactions with anything but the sample, (ii) sample-scattered neutrons, (iii) $\gamma$ rays traveling in the beam and (iv) time-independent background. In Fig.~\ref{fig:bkg} the measured time-of-flight  ($\mathrm{TOF_m}$) spectra, used to estimate these background components, are showed together with the $\mathrm{TOF_m}$ spectrum for the thick $^{155}$Gd sample for comparison.  
The first background component has been evaluated with the empty-sample holder (Sec.~\ref{samples}) in the beam, thus accounting for any beam-related effect not linked to the presence of a sample. The second source of background is attributable to $\gamma$ rays originating from sample-scattered neutrons thermalized and captured in the surrounding materials. It was evaluated with a measurement of the lead sample placed in the beam. To estimate this component the counts of the empty-sample measurement, normalized to the same neutron intensity, were subtracted to the Pb measurement and a correction factor $R_n$ was applied to the resulting quantity. $R_n$ is the ratio of the neutron scattering yield of Gd and Pb sample. The third  background component, mainly 2.2~MeV and 0.48~MeV in-beam $\gamma$~rays from neutron capture in the Hydrogen and Boron of the moderator, respectively, was estimated by a measurement with the natural lead sample. This kind of background starts to contribute in the energy region above 300 eV ($\mathrm{TOF_m}\lesssim 7.7\times10^5$ ns). Its time distribution results from the combination of the neutron slowing-down process in the moderator and the flight path length. 
The fourth background component, related to ambient radioactivity and activation of the materials inside the experimental area was estimated with a beam-off measurement. Since the products of $^{155}$Gd(n,$\gamma$) and $^{157}$Gd(n,$\gamma$) are the stable $^{156}$Gd and $^{158}$Gd isotopes, the background attributable to the activation of the sample is negligible. 

\begin{figure}
\includegraphics[width=0.48\textwidth]{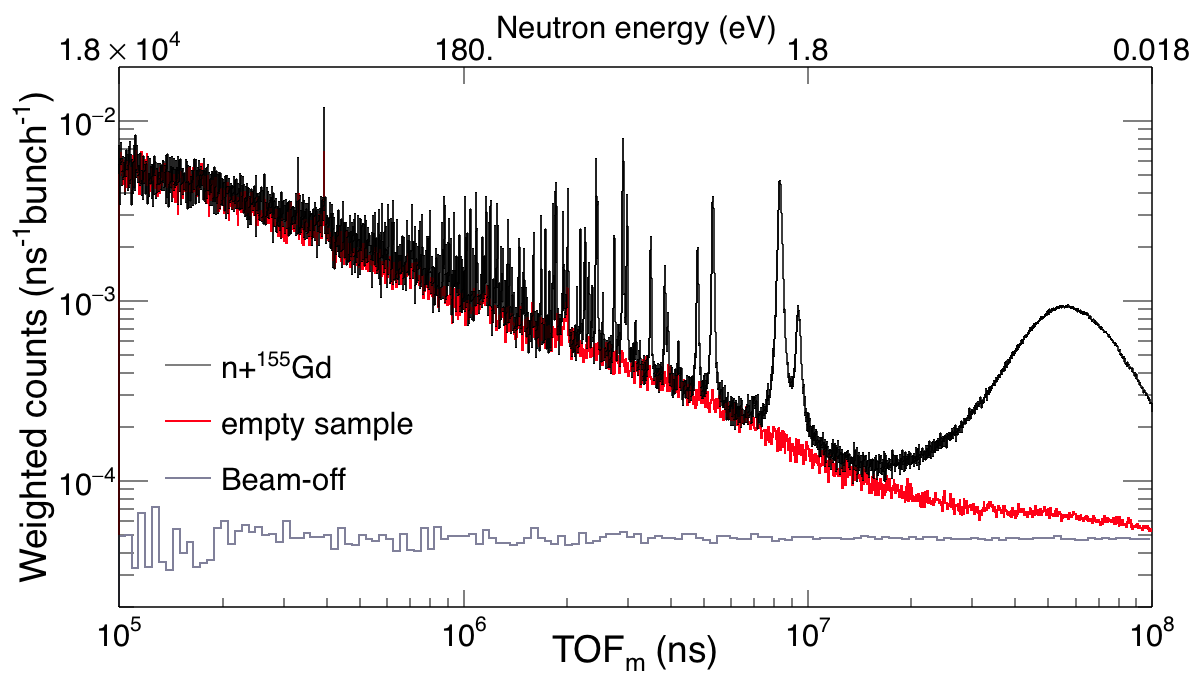}
\caption{(Color online) Weighted C$_6$D$_6$ time-of-flight spectrum of the thin $^{155}$Gd sample, together with background measurements. \label{fig:bkgthin}}
\end{figure}
Fig.~\ref{fig:bkgthin} shows the C$_6$D$_6$ background measurements compared with the signals resulting from the measurement with the thin $^{155}$Gd sample. In this case the neutron background is not shown to highlight the comparison between the Gd and the empty measurement. As expected, the signal-to-background ratio is much less favourable, for instance at 2 eV it is about 200 and 30 for thick and thin samples, respectively. The figure also shows the expected matching between resonance valleys and the empty-sample measurement.  

In summary, the empty sample-measurement satisfactorily represents the background level in the energy range of interest. A similar evaluation was repeated for the thin and thick  $^{157}$Gd(n,$\gamma$) measurement and resulted in the same conclusion. It is worth mentioning that at thermal-neutron energy ($\mathrm{TOF_m}\approx 85\times10^{6}$ns) the signal-to-background ratio for the thin gadolinium samples was 10 and 7.6 for $^{155}$Gd and $^{157}$Gd, respectively.

\subsection{Time-of-flight to energy calibration\label{T2E}}
\begin{figure}
\includegraphics[width=0.48\textwidth]{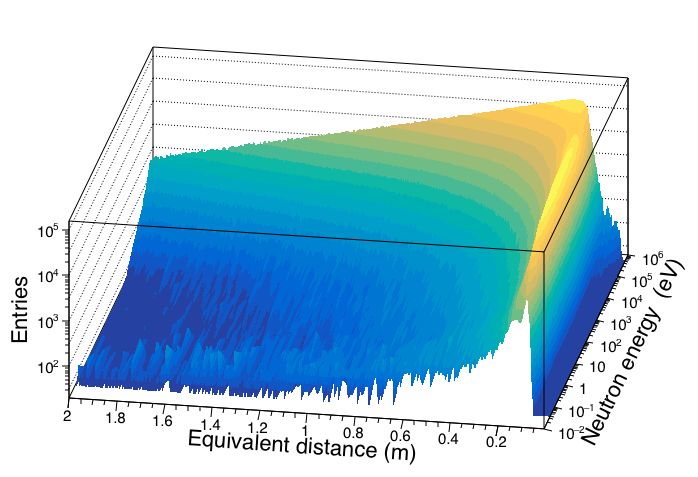}
\caption{(Color online) Distribution of $\lambda(E_n)$ as a function of neutron energy. \label{fig:lambda}}
\end{figure}

The kinetic neutron energy was calculated from the velocity of the neutrons. The latter quantity cannot be directly calculated from the ratio of  the measured $\mathrm{TOF_m}$ and the geometrical flight-path length $L_0$, because $\mathrm{TOF_m}$ also depends on the moderation time  (i.~e. the time spent by the neutron inside the spallation target assembly). This case was discussed in more details in Refs.~\cite{nTOF} and \cite{G4}, here we only remind the reader that it is customary to express the distribution of the moderation time in terms of an equivalent distance $\lambda(E_n)$, which depends on the neutron energy.  Consequently, the effective flight path $L(E_n)$ is the sum of a time independent term $L_0$ and a time dependent term, which is given by the mean of $\lambda(E_n)$ distribution: $L(E_n)=L_0+\langle \lambda(E_n)\rangle$. Fig.~\ref{fig:lambda} shows the $\lambda(E_n)$ distribution obtained by Monte Carlo simulation~\cite{FLUKA1,FLUKA2}, it is worth noticing that the mean of the distribution varies only slightly  with neutron energy around the value of 19~cm.  Therefore the kinetic  energy was extracted with a recursive procedure which converged after few iterations. The value of $L_0=183.92(8)$ m resulted from a minimization procedure adopting the well-known low-energy resonances of $^{197}$Au retrieved from JEFF-3.3 evaluation~\cite{SIRAKOV}. This geometrical (i.~e. energy independent) value, obtained with the least square adjustment is consistent with the nominal value of 183.94 m.
\subsection{Normalization and Beam Interception Factor\label{BIF}}
The normalization factor ($N$ in Eq.~\ref{eq:yield}) groups together the correction factors independent of neutron energy:  efficiency related to the solid angle of the detectors, fraction of the neutron beam intercepting the gadolinium sample (0.68 being the nominal value~\cite{nTOF}) and absolute value of the neutron flux. This normalization coefficient was obtained by the saturated resonance technique~\cite{NDS}, applied to the 4.9 eV resonance in $n+{}^{197}$Au. The Au capture yield was analyzed with the R-Matrix code {\sc SAMMY} ~\cite{SAMMY} and the value of the normalization was extracted with an uncertainty attributable to counting statistic of less than 0.1\%. The impact of the systematic effects related to the difference of the electromagnetic cascade for $^{197}$Au(n,$\gamma$) and $^{155,157}$Gd(n,$\gamma$) is discussed in Sec.\ref{sec:QA}. 

The beam interception factor (i.~e. the fraction of the neutron beam intercepting the sample) can be considered constant within less than 1.5\% variation in the energy region between $E_n=1$ eV and 100 keV~\cite{nTOF}. As thoroughly discussed in Refs.~\cite{nTOF,G4}, the beam profile has a Gaussian distribution with standard deviation of about 6 mm, determined mainly by the collimation system. Below  1 eV, Monte Carlo simulations and experiments have demonstrated that a correction factor, $f_{BIF}$ in Eq.~\ref{eq:yield}, is required for taking into account the modification of the spatial distribution of the beam profile. 
Unfortunately, the correction is extremely sensitive to small changes in the collimation system, which cannot be fully controlled and therefore implemented in MC simulations. In addition also gravitational force plays a sizable role for very low-energy neutrons, because the vertical displacement is 3.5 cm for neutrons of 25 meV, after traveling for 185 m. 
Therefore from the Monte Carlo simulation only qualitative information could be drawn: the beam profile becomes larger and asymmetric as the neutron energy decreases and the correction can be as high as 20\%.

\begin{figure}
\includegraphics[width=0.48\textwidth]{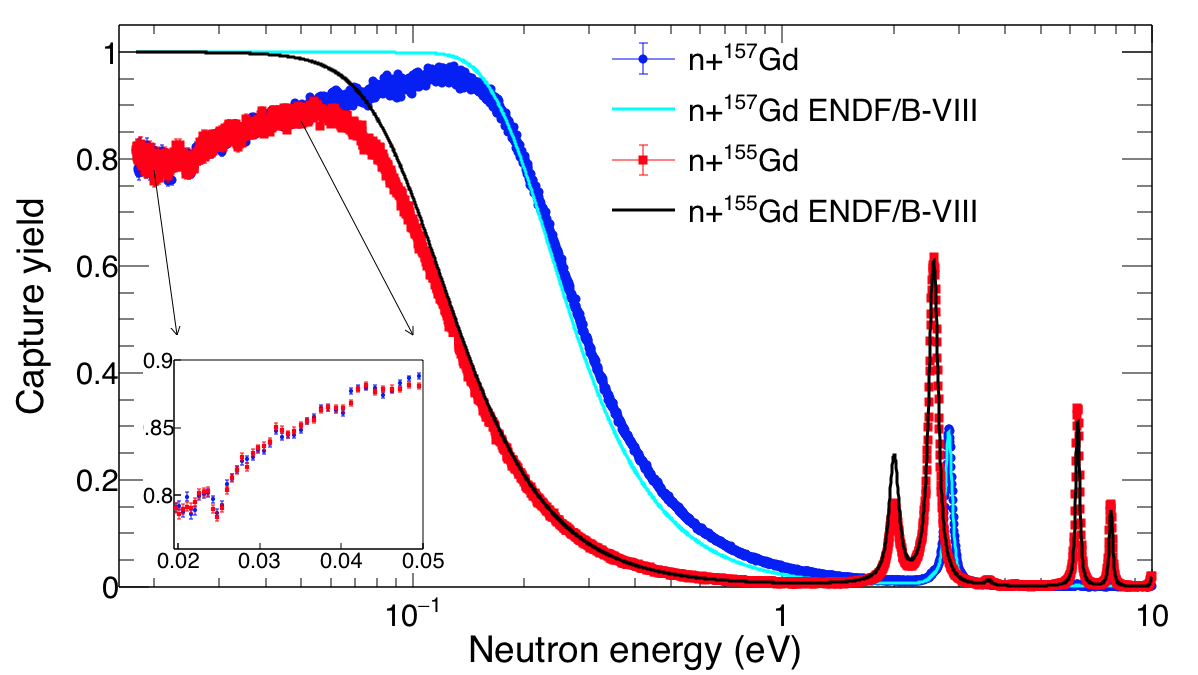}
\caption{(Color online) Capture yield of the thick gadolinium samples measured in this work and comparison with the expected capture yield calculated on the basis of the cross sections in ENDF/B-VIII.0 library. The region between $E_n=0.02$ and 0.05 eV, linked to the correction for the variation of the beam interception factor, is shown in the inset. \label{fig:BIF2}}
\end{figure}

In the present analysis,  an empirical method was used for the correction of the beam interception factor. It is similar to the saturated resonance technique (i.~e. the product of the areal density and cross section is high enough for all incident neutrons to interact with the sample), 
the expected capture yield can be expressed as: 
\begin{equation}\label{eq:yield_teo}
Y(E_n)=(1-e^{-n\sigma_{tot}(E_n)})\frac{\sigma_\gamma(E_n)}{\sigma_{tot}(E_n)}+Y_m,
\end{equation} 
 where $n$ is the areal density of the sample (reported in Table~\ref{table:sample}), $\sigma_{tot}$ is the total cross section, $\sigma_\gamma$  is the capture cross section and $Y_m$ accounts for the contribution of capture events following at least one neutron scattering in the sample. In the case of the thick $^{155}$Gd and $^{157}$Gd samples, $n\sigma_{tot}(E_n)$  is so high that the calculated transmission of neutrons through the samples is less than $10^{-3}$, for neutron energies below 0.07 and 0.1 eV, respectively. Considering also that the ratio of elastic to capture cross section is less than $10^{-2}$ in this energy region, the capture yield for thick gadolinium samples is consequently expected to be $Y=1$. Any departure of the measured capture yield from unity was ascribed to a variation of the beam interception factor, as illustrated in Fig.~\ref{fig:BIF2}, where the experimental $^{155}$Gd(n,$\gamma$) and $^{157}$Gd(n,$\gamma$) capture yields are compared to their expected values based on the resonance parameters in the ENDF/B-VIII.0 evaluation.  In the inset of the figure, the energy region where the empirical  $f_{BIF}$ was extracted is highlighted. The two sets of data are very similar, confirming the presence of a common effect. 

The empirical correction factor was used to correct the capture yields of the thin gadolinium samples and the gold one, as discussed in the next section.
\subsection{Quality assessment and discussion on uncertainties\label{sec:QA}}
As mentioned above, one of the aims of this measurement was to estimate the cross section at thermal energy for $^{155}$Gd and $^{155}$Gd. In this region the uncorrelated uncertainties have a minor role, since the involved cross sections are very high and consequently the counting statistics is not the issue. On the contrary, the correlated uncertainties dominate the total uncertainty. They come from the normalization, PHWT, background determination and subtraction, sample characterization and neutron flux shape. In addition, in the energy region below 1 eV the uncertainty due to correction for the beam interception factor should also be considered. In Table~\ref{tab:unc} the different contributions are listed together with the total uncertainty for both gadolinium isotopes in the thermal region and in the resonance region. Hereafter each component is discussed separately.
 
The uncertainty related to the normalization depends on the differences between the electromagnetic cascades in $^{197}$Au(n,$\gamma$) and in $^{155,157}$Gd(n,$\gamma$). One effect is attributable to the electron conversion and the other to the missing $\gamma$ rays because of the discrimination level in the detectors. To quantify such a bias the same method as in Ref.~\cite{MASSIMI} was applied. The count losses attributable to the detector threshold were estimated by means of simulated cascades (as described above, in Sec.~\ref{WF}). The weighted contribution of $\gamma$ rays with an energy below the threshold and of electrons is 1.0\% and 0.8\%, respectively, for $^{198}$Au. Whereas for both $^{156}$Gd and $^{158}$Gd the weighted contributions are similar and their value is 0.4\% for missing $\gamma$ rays  and 0.2\% for conversion electrons. Therefore the bias in the normalization due to the detector threshold is 0.6\% for missing $\gamma$ rays  and 0.6\% for conversion electrons. Since these are model-dependent corrections, which depends also in the detector performances, they were used only to study the uncertainty on the normalization which was estimated to be less than 1.5\%. 


The uncertainty related to the weighting function 
was investigated by repeating
 the data analysis by using: i) linear and quadratic amplitude to deposited energy calibration;  ii) detector threshold of 150, 175 and 200 keV and corresponding weighting functions; (iii) 7 different weighting functions calculated with an exponential attenuation in the direction of the neutron beam according to different values of $n\sigma_{tot}$. In any case the ratio between the experimental yields never changed by more than 1.5\%. Moreover, systematic effects due to the positioning of the sample with respect to the detection system and the neutron beam were minimized. In particular samples were centered using a micrometric positioning system based on a jig and a hollow metallic cylinder aligned with the Al annular frame. 

The component related to the background subtraction propagates to the total uncertainty according to the signal-to-background ratio.
From measurements using different empty samples, and from the comparison of the TOF spectra measured with the empty sample and the thin gadolinium samples (between resonances), we deduced an uncertainty of 10\%.
 Therefore, at thermal neutron energy the uncertainty attributable to the background subtraction is 1.4\% for $^{155}$Gd and 1.0\% for $^{157}$Gd (the signal-to-background ratio is 7.6 and 10, respectively), whereas in the resonance region it depends on the resonance strength.  

For the estimation of the uncertainty of the shape of the neutron flux, we adopted an uncertainty of 1\% as discussed in~\cite{BARB,nTOF}. This value corresponds to the uncertainty in the region around 4.9 eV, where the normalization was extracted. 

\begin{figure}
\includegraphics[width=0.48\textwidth]{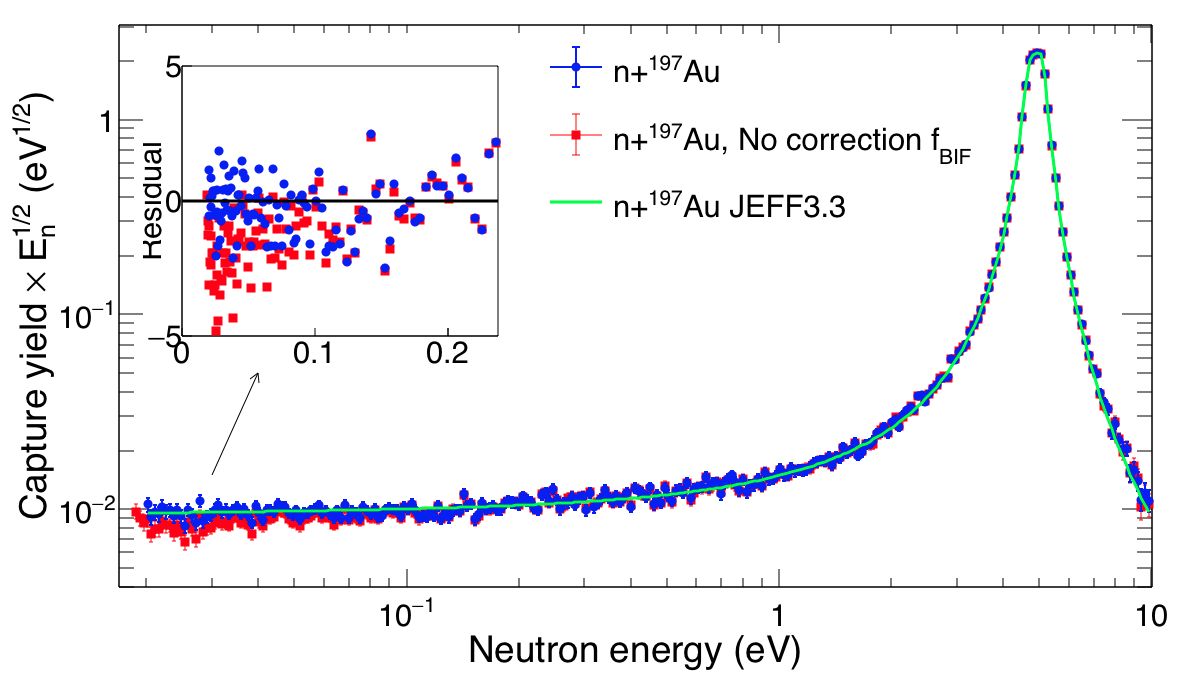}
\caption{(Color online)  $^{197}$Au(n,$\gamma$) capture yields (multiplied by $E^{1/2}$) with and without the correction for the variation of the beam interception factor and comparison with the expected capture yield calculated on the basis of the cross sections in JEFF-3.3 library. \label{fig:au}}
\end{figure}

The uncertainty on the correction of the beam interception factor was estimated by analyzing the capture yield of $^{197}$Au(n,$\gamma$), since its cross section is a standard at thermal energy~\cite{CARL}. In addition the cross section near thermal energy is characterized by the $1/v$ behavior and therefore the knowledge of the capture yield at thermal energy constraints the shape of the capture yield at higher energies as well. In Fig.~\ref{fig:au} two experimental $^{197}$Au(n,$\gamma$) capture yields are showed together with the expected capture yield based on the JEFF-3.3 evaluation. The two data sets differ by applying the correction for the beam interception factor. It is evident that near thermal energy (highlighted in the inset) the expected capture yield is reproduced within 1-2\% when the correction is applied. In particular, the average deviation between the present data and the expected capture yield in the region $0.02 \le E_n \le 0.1$ eV is 0.985 with a root mean square of 0.06. From this evidence an uncertainty of 2\% was estimated for the beam interception factor.

Very thin samples can suffer from inhomogeneity because of the preparation procedure. The provider claimed an uncertainty in the uniformity better than 10\%. 
To constrain this possibly large uncertainty, we have compared the results of the resonance shape analysis for strong resonances as observed in the capture yields of thin and thick samples, see Sec.~\ref{Sec_4}. For both pairs of samples the results were consistent within 1\%, hence the uncertainty in the uniformity can be accounted for as a part of the Sample mass uniformity. This comparison was also the confirmation that the uncertainties summarized in Table~\ref{tab:unc} can be considered as the full uncertainty of the present measurements.

\begin{table*}
\caption{Summary of the correlated uncertainties in the $^{155}$Gd(n,$\gamma$) and $^{157}$Gd(n,$\gamma$) cross section measurements.\label{tab:unc}}
\begin{ruledtabular}
\begin{tabular}{lcccc}
Source of&  \multicolumn{2}{c}{$^{155}$Gd(n,$\gamma$)} & \multicolumn{2}{c}{$^{157}$Gd(n,$\gamma$)} \\
uncertainty & near thermal & resonance region& near thermal & resonance region\\
\hline
PHWT            &1.5\%&1.5\%&1.5\%&1.5\% \\
Normalization &1.5\%&1.5\%&1.5\%&1.5\% \\ 
Background   &1.4\%&$\approx1$\%&1.0\%&$\approx1$\% \\ 
Sample mass &1.0\%&$<0.1$\%&2.1\%&$<0.1$\% \\
BIF                  &2.0\%& &2.0\%& \\
Flux               &1.0\%&1.0\%&1.0\%&1.0\% \\
\hline
{\bf Total} &{\bf 3.5\%}&{\bf 2.5\%}&{\bf 3.9\%}&{\bf 2.5\%}\\
 \end{tabular}
 \end{ruledtabular}
 \end{table*}
\section{Resonance shape Analysis \label{Sec_4}}
The  capture yields were analyzed with the R-Matrix analysis code {\sc SAMMY}, using the Reich-Moore approximation. Corrections for experimental conditions such as Doppler and experimental broadening, self-shielding and multiple neutron interactions in the sample (i.~e. multiple scattering) were taken into account by the code. In particular, the response of the spectrometer (showed in Fig.~\ref{fig:lambda} ) was implemented in {\sc SAMMY} by using the user-defined resolution function option: i.~e. a numerical description derived from a Monte Carlo simulations. The thermal motion of gadolinium atoms inside the sample was taken into account by means of the free-gas model with a temperature of 296 K, as monitored during the experiment.

The resonance parameters and the scattering radius from the ENDF/B-VIII.0 library were adopted as the initial values of a fitting procedure. The scattering radius as well as the spin and parity of the resonances were not changed, because the capture data are not sensitive enough to these quantities. In the analysis, the resonance energy and both $\Gamma_n$ and $\Gamma_\gamma$ were varied because the improvement in the $\chi^2$ value of the fit was substantial with respect to the case where only one parameter (either $\Gamma_n$ or $\Gamma_\gamma$) was allowed to vary and the other was fixed to the ENDF/B-VIII value. Since the spin assignments in the evaluations are sometimes inconsistent and do not take into account recent results~\cite{BARA}, $g\Gamma_n$, $g$ being the statistical spin factor, is reported in this work since its value is independent of the spin of the resonance.

For energies below 0.5 eV, only the data obtained with the thin samples were used. A simultaneous resonance shape analysis of both data obtained with thin and thick sample was performed up to 5 eV. Above this energy, only the data obtained with the thick samples were used. Nevertheless, as already mentioned above, few strong resonances in the energy region up to 60 eV were used to cross-check the capture data obtained with the thin samples. 

The results of the resonance shape analysis were used to reconstruct the cross section and in particular to evaluate the thermal cross section $\sigma_0=\sigma_\gamma(E_0 )$ for $E_0=0.0253$ eV. In addition, the cross section reconstructed using the resonance parameters from this work has been convoluted with a Maxwellian neutron energy distribution to obtain the so called Maxwellian averaged cross section: 
\begin{equation}\label{eq:MACS}
\frac{\int{\sigma_\gamma(E_n)\frac{E_n}{E_0}e^{-E_n/E_0}}dE_n}{ \int{\frac{E_n}{E_0}e^{-E_n/E_0}}dE_n}.
\end{equation}
  The ratio of  the latter quantity for thermal energy $E_0=0.0253$ eV to the thermal cross section, also referred to as the Westcott factor~\cite{MUGH}, was also calculated. It allowed us to evaluate the non-$1/v$ behaviour of the capture cross section (i.~e. the Westcott factor significantly different from unity).

The resolved resonance regions (RRR) in  nuclear data libraries such as ENDF/B-VIII.0, JEFF-3.3 and JENDL-4.0 are limited to the energy region below 300 eV for n+$^{157}$Gd and below 180 eV for n+$^{155}$Gd. The present data clearly show structures well above these energies (see Fig.\ref{fig:GdURR} in Appendix~\ref{App:newres}). These structures have been analyzed assuming they are $s$-wave resonances with an average $\Gamma_\gamma$ deduced from the resonances in the RRR. Their energy and capture kernel, defined as $g\Gamma_\gamma\Gamma_n/(\Gamma_\gamma+\Gamma_n)$ are reported in Appendix~\ref{App:newres}. In order to extend the resolved resonance region to higher energies, it would be necessary to perform a transmission experiment on the same thick samples. 
\subsection{n+$^{155}$Gd\label{sec155}}
The capture cross section of $^{155}$Gd at thermal energy does not vary significantly among libraries. It ranges from  60.735 to 60.890  kb. From the present data a slightly higher but consistent thermal cross section was deduced, $\sigma_0=62.2(2.2)$ kb. The resulting Westcott factor is $0.86(4)$. The weighted mean of the present result combined with the cross section values by M{\o}ller~\cite{MOLLER} and Ohno~\cite{OHNO} provides $\sigma_0=60.2(4)$ kb, which is consistent with the value reported by Leinweber~\cite{LEINW}. 

In the resolved resonance region, differences are present in evaluated nuclear data files. Moreover, two  time-of-flight measurements present inconsistencies. In particular, for a number of resonances, the measurement of Leinweber and collaborators~\cite{LEINW} sizably disagrees with the ENDF/B-VIII.0, while the measurement by Baramsai and collaborators~\cite{BARA} tends to confirm the resonance parameters in ENDF/B-VIII.0.

\begin{figure}
\includegraphics[width=0.44\textwidth]{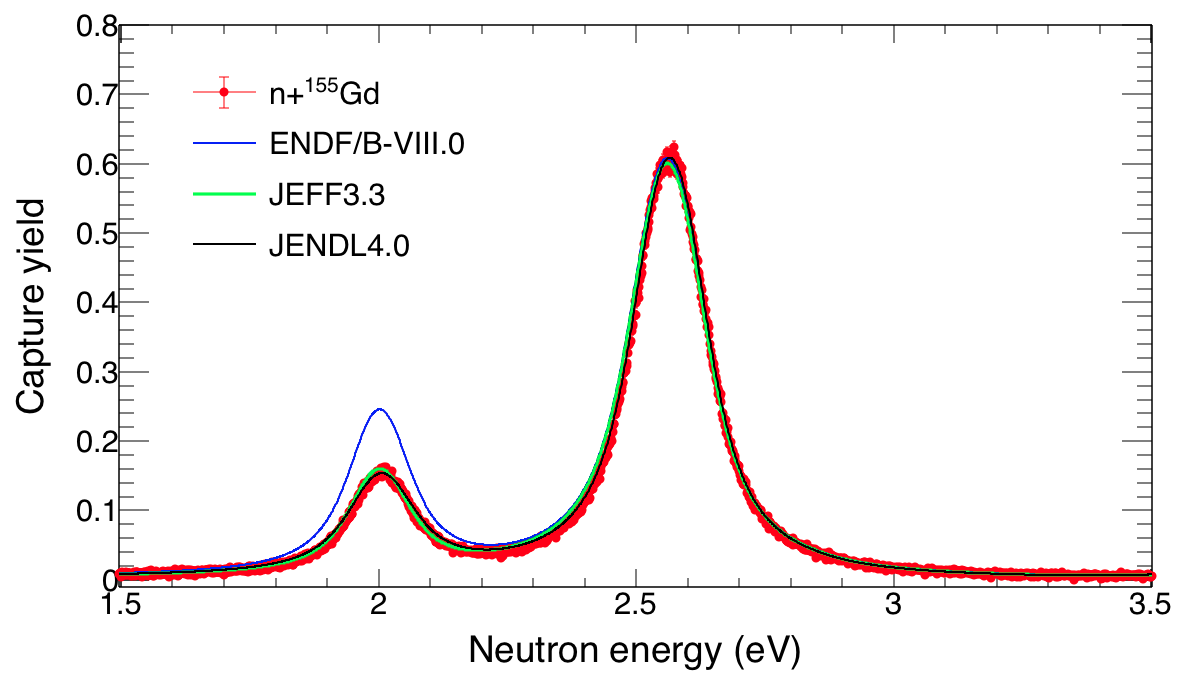}
\includegraphics[width=0.48\textwidth]{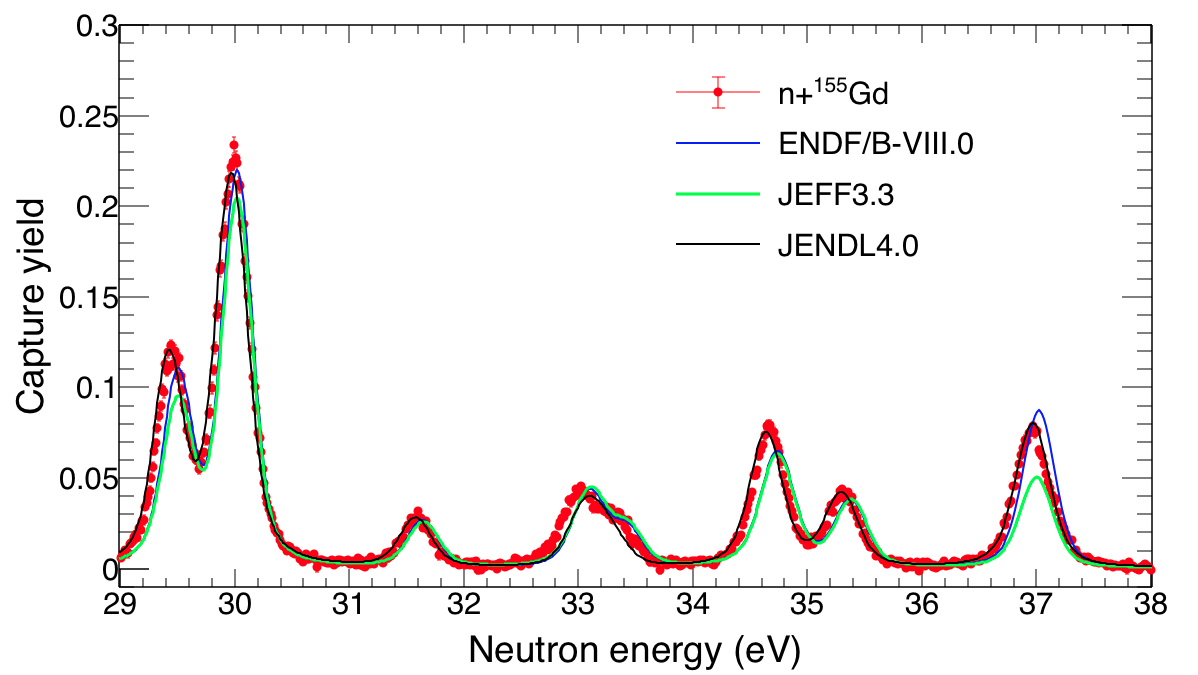}
\includegraphics[width=0.48\textwidth]{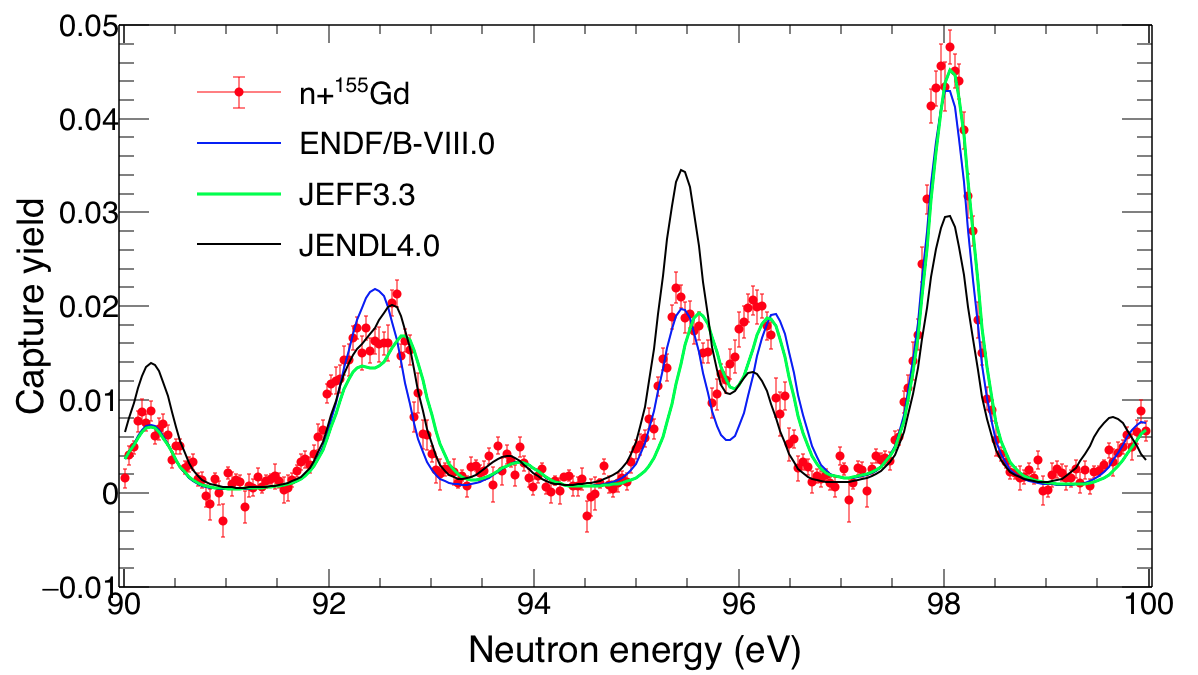}
\caption{(Color online)  $^{155}$Gd(n,$\gamma$) capture yield from the present work compared to the expected capture yields, calculated on the basis of the cross sections in ENDF/B-VIII.0, JEFF-3.3 and JENDL-4.0 nuclear data libraries. \label{fig:Gd5}}
\end{figure}

Examples of some of the largest differences between the present data and the evaluations are shown in Fig.~\ref{fig:Gd5}. For instance the kernel of the resonance at 2.0 eV is about 50\% lower than the value calculated from ENDF/B-VIII.0. Large deviations with respect to JENDL-4.0 are present as well, as for example in the case of the resonances at 95.7 eV and 98.3 eV.  
The bottom panel in Fig.~\ref{fig:Gd5} confirms the good energy resolution of the present data, able to resolve some doublets, such as the structures at $E_n=33$, 93 or 96 eV. 
It is worth noticing that the resonances at 22.45, 116.9 and 138.0 eV in Ref.~\cite{BARA} could be attributable to multiple scattering in the nearby resonances rather than genuine resonance as quoted in ENDF/B-VIII (see for instance Fig. 21 in Ref.~\cite{ORO} about this possible effect). The presence of small structures at 43.43, 54.81, 62.12, 68.98 and 76.19 eV reported in JEFF-3.3 and 17.81 and 68.81 eV reported by Baramsai and collaborators~\cite{BARA},  cannot be excluded on the basis of present data since they are overwhelmed by the background. Finally, the resonance reported by Leinweber and collaborators at $E_n=131.7$ eV is also observed in the present measurement.   

The results of the resonance shape analysis are summarized in Table~\ref{tab:res155}. The correlation coefficient between partial widths $\rho(\Gamma_\gamma, \Gamma_n)$ resulting from the {\sc SAMMY} fit is also reported.
 \begin{longtable}{@{\extracolsep{\fill}}lcccc@{}}
\caption{Resonances in $^{155}$Gd(n,$\gamma$). Uncertainties are from the fit. \label{tab:res155}}\\
\hline 
\hline
Energy&$\Gamma_\gamma$ & g$\Gamma_n$ & $\rho$($\Gamma_\gamma$,$\Gamma_n$)&capure kernel\\
(eV)  & (meV)& (meV) & &(meV)\\ 
\hline
\endfirsthead
\multicolumn{5}{c}{ \tablename\ \thetable{}  (Continued) } \\ \hline
Energy&$\Gamma_\gamma$ & $g\Gamma_n$ & $\rho$($\Gamma_\gamma$,$\Gamma_n$)&capture kernel\\
(eV)  & (meV)& (meV) & &(meV)\\ 
\hline
\endhead
\hline \multicolumn{5}{c}{{Continued}} \\ 
\endfoot
\hline \hline
\endlastfoot
0.0268(0)	&	104.57(8)  &	0.0644(5)	&	0.48	&	0.0643(5)\\
2.0128(2)	&	111.5(6)  &	0.1350(4)&	0.46	&	0.1348(4)\\
2.5730(1)	&	103.1(2)  &	1.037(1)&	$-0.17$	&	1.021(1)\\
3.618(2)	&	123(6)	&	0.0141(4)&	0.64	&	0.0141(4)\\
6.3062(2)	&	103.9(5)&	1.331(3)&	0.07	&	1.304(3)\\
7.7490(4)	&	102.4(8)&	0.697(3)&	0.33	&	0.689(3)\\
10.000(2)	&	99(4)	&	0.105(2)&	0.60	&	0.105(2)\\
11.507(1)	&	104(4)	&	0.217(3)&	0.61	&	0.216(3)\\
11.9729(7)&	107(2)	&	0.650(4)&	0.52	&	0.644(4)\\
14.4851(6)&	102(5)	&	1.233(6)&	0.28	&	1.195(6)\\
17.733(2)	&	92(5)	&	0.234(5)&	0.53	&	0.233(5)\\
19.8790(6)&	107(1)	&	2.76(1)	&	0.25	&	2.65(1)	\\
20.9902(4)&	120.9(1)&	7.24(2)	&	$-0.12$	&	6.61(2)	\\
23.602(1)	&	137(7)	&	1.87(1)	&	0.50	&	1.84(1)	\\
27.519(3)	&	96(6)	&	0.403(9)&	0.54	&	0.399(8)\\
29.528(1)	&	112(3)	&	2.95(3)	&	0.57	&	2.83(3)	\\
30.0702(7)&	109(2)	&	6.89(3)	&	0.27	&	6.26(3)	\\
31.674(3)	&	93(6)	&	0.66(1)	&	0.47	&	0.66(1)	\\
33.047(3)	&	121(8)	&	0.89(2)	&	0.64	&	0.89(2)	\\
33.464(5)	&	103(9)	&	0.54(2)	&	0.48	&	0.52(2)	\\
34.758(1)	&	114(4)	&	2.46(2)	&	0.38	&	2.34(2)	\\
35.408(2)	&	107(6)	&	1.21(2)	&	0.57	&	1.19(2)	\\
37.067(2)	&	139(4)	&	2.90(2)	&	0.31	&	2.75(2)	\\
38.937(2)	&	99(7)	&	0.63(1)	&	0.46	&	0.62(1)	\\
43.868(1)	&	121(8)	&	6.52(5)	&	$-0.15$	&	6.06(4)	\\
46.006(3)	&	121(8)	&	1.32(3)	&	0.51	&	1.30(3)	\\
46.806(2)	&	98(4)	&	3.44(3)	&	0.37	&	3.26(3)	\\
47.64(1)	&	98(10)	&	0.20(1)	&	0.20	&	0.20(1)	\\
51.290(2)	&	139(4)	&	7.58(6)	&	0.13	&	6.61(6)	\\
52.041(1)	&	101(4)	&	7.74(6)	&	0.01	&	6.42(6)	\\
52.918(7)	&	124(11)	&	0.843)	&	0.48	&	0.83(3)	\\
53.639(2)	&	99(4)	&	5.09(5)	&	0.17	&	4.70(4)	\\
56.130(4)	&	110(8)	&	1.36(3)	&	0.38	&	1.33(3)	\\
59.321(2)	&	112(5)	&	4.20(5)	&	0.21	&	3.97(4)	\\
62.751(2)	&	126(5)	&	4.82(5)	&	0.21	&	4.54(5)	\\
64.06(3)	&	124(12)	&	0.14(1)	&	0.06	&	0.14(1)	\\
65.17(2)	&	111(18)	&	0.35(2)	&	0.25	&	0.35(2)	\\
69.459(3)	&	123(7)	&	3.77(6)	&	0.35	&	3.74(5)	\\
76.835(8)	&	103(10)	&	0.95(3)	&	0.32	&	0.93(3)	\\
77.60(2)	&	112(11)	&	0.37(2)	&	0.15	&	0.37(2)	\\
78.761(7)	&	124(10)	&	2.64(5)	&	0.32	&	2.56(5)	\\
78.9(2)	&	120(12)	&	0.10(1)	&	0.10	&	0.095(9)\\
80.70(1)	&	111(10)	&	0.88(3)	&	0.14	&	0.88(3)	\\
83.980(4)	&	109(7)	&	3.67(7)	&	$-0.05$	&	3.67(6)	\\
84.916(9)	&	113(10)	&	1.27(4)	&	0.32	&	1.27(4)	\\
90.53(1)	&	111(11)	&	0.75(3)	&	0.21	&	0.74(3)	\\
92.44(1)	&	105(10)	&	1.48(7)	&	0.26	&	1.45(7)	\\
92.89(1)	&	96(9)	&	1.80(8)	&	0.29	&	1.75(7)	\\
93.94(3)	&	111(11)	&	0.31(2)	&	0.07	&	0.31(2)	\\
95.710(8)	&	117(11)	&	2.37(7)	&	0.37	&	2.30(7)	\\
96.403(9)	&	106(10)	&	2.41(8)	&	0.25	&	2.27(7)	\\
98.302(4)	&	98(7)	&	7.4(1)	&	$-0.52$	&	7.40(9)	\\
100.21(2)	&	130(13)	&	0.74(4)	&	0.14	&	0.73(4)	\\
101.35(1)	&	120(12)	&	1.38(5)	&	0.33	&	1.35(5)	\\
101.99(2)	&	95(10)	&	0.83(4)	&	0.16	&	0.81(4)	\\
104.413(7)	&	118(11)	&	3.27(8)	&	0.11	&	3.05(8)	\\
105.942(8)	&	120(11)	&	2.34(7)	&	0.16	&	2.22(7)	\\
107.118(6)	&	119(10)	&	3.96(8)	&	0.32	&	3.76(8)	\\
109.55(1)	&	112(11)	&	1.57(6)	&	0.16	&	1.51(6)	\\
112.389(5)	&	113(9)	&	5.5(1)	&	0.13	&	5.08(9)	\\
113.822(3)	&	128(7)	&	11.1(1)	&	$-0.15$	&	9.8(1)	\\
116.541(5)	&	127(9)	&	7.7(1)	&	$-0.19$	&	6.6(1)	\\
118.69(2)	&	107(11)	&	1.12(6)	&	0.10	&	1.10(6)	\\
123.377(4)	&	105(10)	&	16.4(2)	&	$-0.44$	&	16.4(2)	\\
124.448(8)	&	106(8)	&	4.2(1)	&	0.10	&	4.2(1)	\\
126.102(5)	&	127(10)	&	9.1(2)	&	$-0.26$	&	7.6(1)	\\
128.55(6)	&	104(10)	&	0.30(3)	&	0.02	&	0.33(3)	\\
129.73(4)	&	109(11)	&	0.80(7)	&	0.06	&	0.80(6)	\\
130.877(6)	&	121(10)	&	13.2(2)	&	0.29	&	13.2(2)	\\
133.04(2)	&	105(11)	&	1.47(8)	&	0.11	&	1.42(7)	\\
133.88(2)	&	99(10)	&	1.62(8)	&	0.12	&	1.57(8)	\\
134.75(5)	&	99(10)	&	0.52(4)	&	0.05	&	0.51(4)	\\
137.809(8)	&	117(10)	&	5.0(2)	&	0.29	&	5.0(2)	\\
140.39(2)	&	99(10)	&	1.34(7)	&	0.11	&	1.30(7)	\\
141.33(5)	&	109(11)	&	0.48(4)	&	0.05	&	0.48(4)	\\
145.63(1)	&	144(13)	&	3.5(1)	&	0.16	&	3.4(1)	\\
147.02(1)	&	131(13)	&	2.9(1)	&	0.22	&	2.8(1)	\\
148.193(9)	&	146(13)	&	5.5(2)	&	0.23	&	5.2(1)	\\
149.484(8)	&	111(10)	&	12.5(4)	&	$-0.28$	&	9.6(3)	\\
150.176(7)	&	124(10)	&	14.8(3)	&	0.03	&	12.4(3)	\\
152.24(1)	&	106(10)	&	3.2(1)	&	0.05	&	3.0(1)	\\
153.71(6)	&	126(13)	&	0.46(4)	&	0.10	&	0.46(4)	\\
156.291(9)	&	114(11)	&	4.7(1)	&	0.10	&	4.4(1)	\\
160.063(9)	&	112(10)	&	5.5(1)	&	0.06	&	5.1(1)	\\
161.616(5)	&	108(9)	&	12.1(2)	&	$-0.37$	&	10.3(2)	\\
168.311(7)	&	98(9)	&	11.9(4)	&	$-0.64$	&	9.0(2)	\\
170.34(1)	&	115(11)	&	5.4(2)	&	0.13	&	5.1(1)	\\
171.30(1)	&	126(12)	&	5.6(2)	&	0.10	&	5.3(2)	\\
173.566(5)	&	122(9)	&	20.7(4)	&	$-0.59$	&	16.3(2)	\\
175.44(4)	&	116(11)	&	1.07(8)	&	0.04	&	1.05(8)	\\
178.01(1)	&	111(11)	&	4.0(2)	&	0.01	&	3.7(1)	\\
180.32(1)	&	112(11)	&	6.0(2)	&	$-0.16$	&	5.2(1)	\\
\hline
 \end{longtable}
   
\begin{figure}
\includegraphics[width=0.48\textwidth]{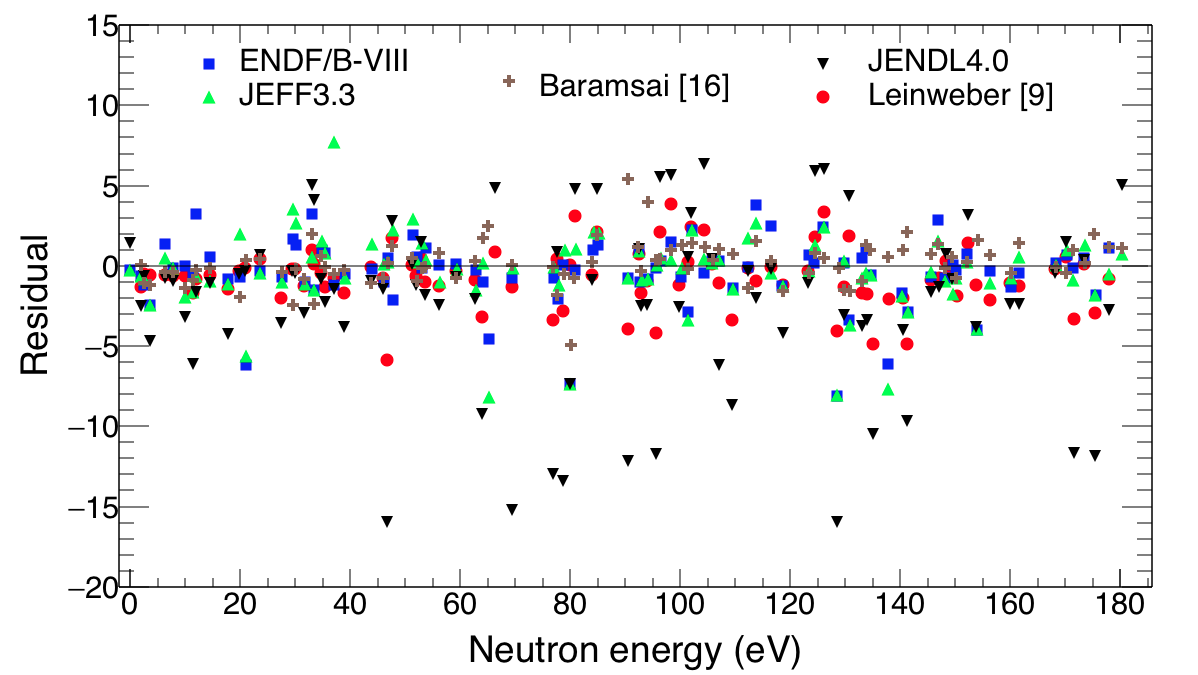}
\caption{(Color online) $^{155}$Gd(n,$\gamma$) residuals between the present resonance kernels and values in literature: ENDF/B-VIII.0, JEFF-3.3 and JENDL-4.0 evaluations and  TOF experiment reported in Ref.~\cite{LEINW} and~\cite{BARA}, as a function of neutron resonance energy. \label{fig:kernel155}}
\end{figure}

A comparison of the kernels from the present analysis to the ones from evaluations, Ref.~\cite{LEINW} and Ref.~\cite{BARA} is reported as a function of resonance energy in Fig.~\ref{fig:kernel155} in terms of residuals (i.~e. difference of our values to the ones in literature, divided by the uncertainty).
On the average a good agreement was found with the ENDF/B-VIII.0 and JEFF-3.3 evaluations, as well as with the resonance parameters by Baramsai {\it et al.}~\cite{BARA}. From the figure it results that the resonance parameters form JENDL-4.0 are not directly taken from Ref.~\cite{LEINW}.
Moreover, the statistical distribution of the ratios of our kernels to the others was Gaussian with mean 0.98, 0.98 and 1.02, respectively. On the contrary, the comparison with JENDL-4.0 and the data from Leinweber and collaborators presents an average deviation of about 8\% and 7\%, respectively.  

The resonances and structures observed in the energy region above the upper limit of evaluations are reported in Tab.~\ref{tab:new55} in Appendix~\ref{App:newres}, together with their capture kernels. 
\subsection{n+$^{157}$Gd\label{sec157}}
\begin{figure}
\includegraphics[width=0.48\textwidth]{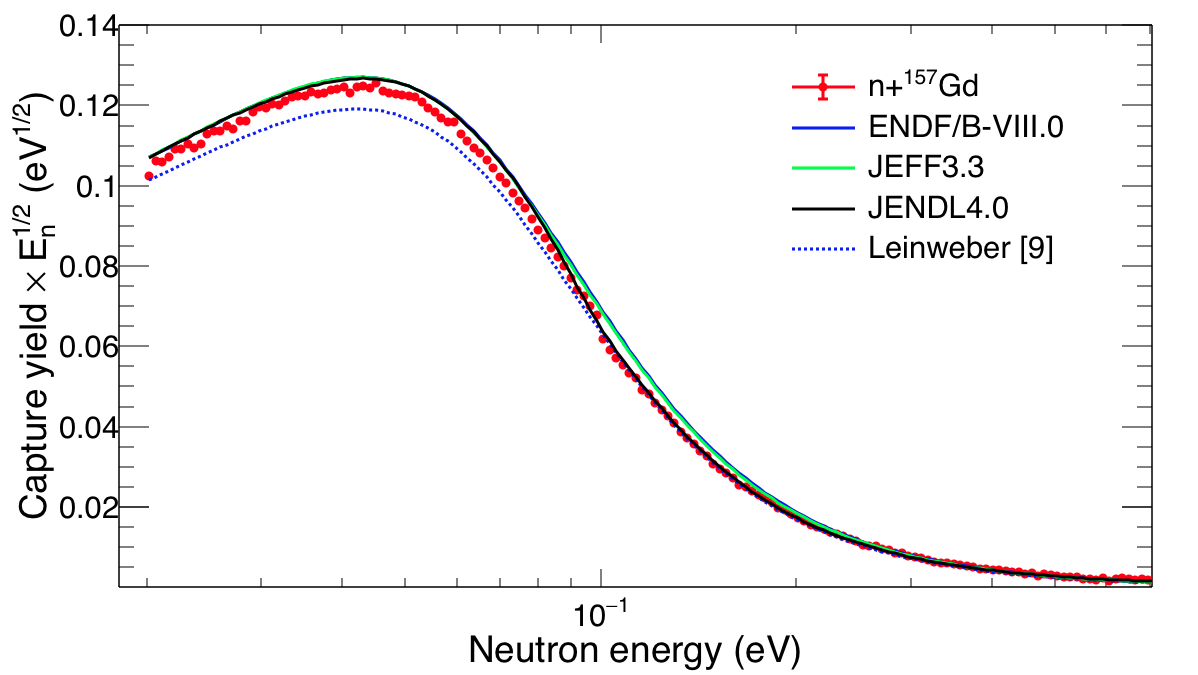}
\caption{(Color online)  $^{157}$Gd(n,$\gamma$) capture yield (multiplied by $E^{1/2}$) from the thin sample compared with the expected capture yields calculated on the basis of the cross sections in ENDF/B-VIII.0, JEFF-3.3 and JENDL-4.0 libraries and resonance parameters in Ref.~\cite{LEINW}.\label{fig:Gd7th}}
\end{figure}

In the region near thermal energy, the three data libraries ENDF/B-VIII.0, JEFF-3.3 and JENDL-4.0 provide similar values of the capture cross section, between 253.2 and 254.5 kb.  In the experiment by Leinweber and collaborators~\cite{LEINW}, a 12\% smaller cross section was deduced. In Figure~\ref{fig:Gd7th} the present capture yield, obtained with the thin sample, is compared with the expected capture yields calculated from the resonance parameters in evaluations and Ref.~\cite{LEINW}.  
The present data settle in between the two groups of expected values. Our estimation of the thermal cross section, deduced from the resonance parameters in Table~\ref{tab:res157} is $\sigma_0=239.8(9.3)$ kb. As in the case of $^{155}$Gd, the Westcott factor sizably deviates from 1, with a value of  $0.89(4)$, some 5\% higher than evaluations. The weighted mean of the present result combined with the cross section values by M{\o}ller~\cite{MOLLER} and Ohno~\cite{OHNO} provides $\sigma_0=252(2)$ kb, which is not consistent with the results of Ref.~\cite{LEINW}.

\begin{figure}
\includegraphics[width=0.48\textwidth]{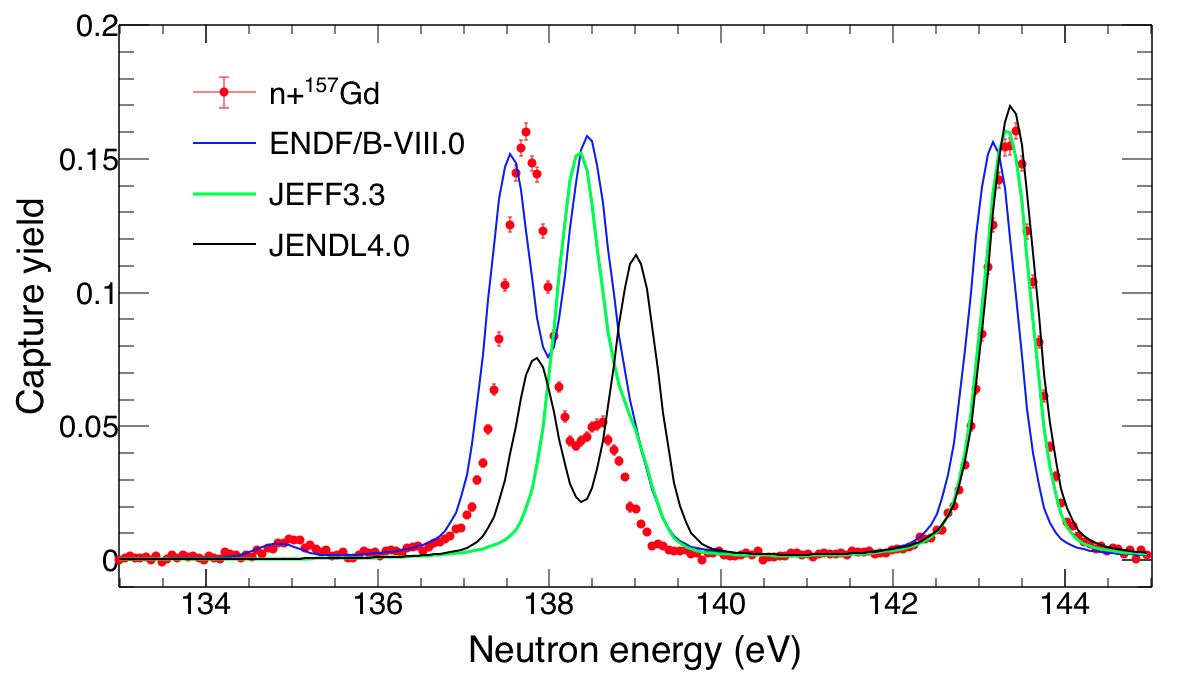}
\includegraphics[width=0.48\textwidth]{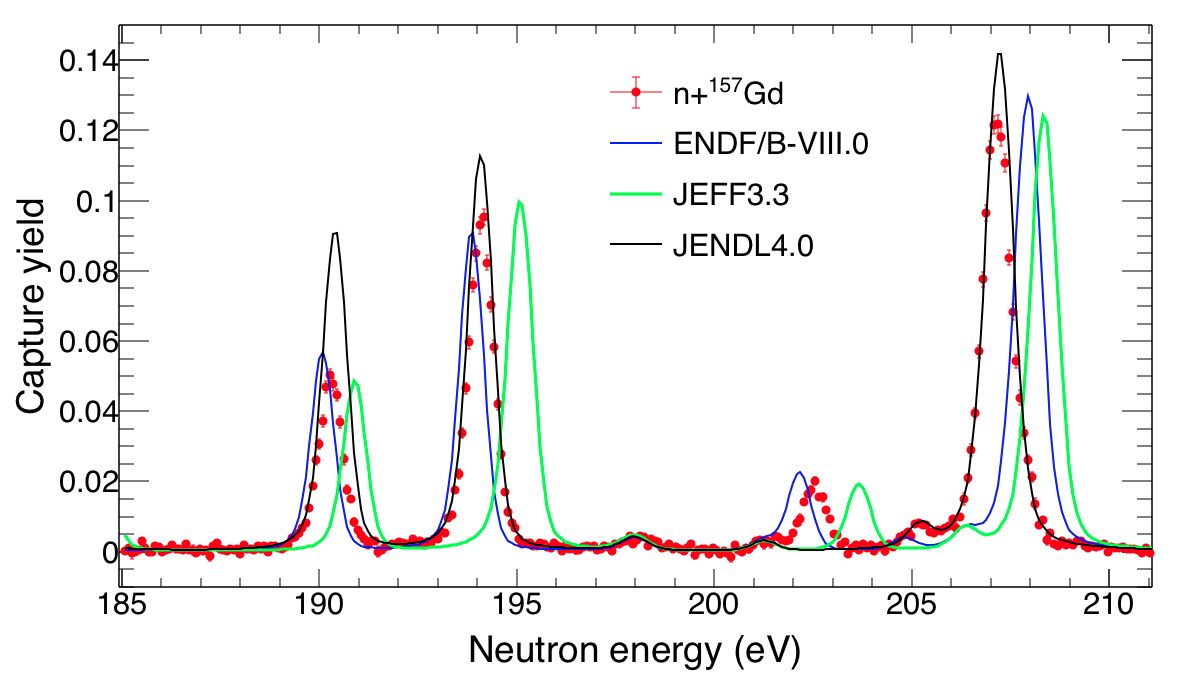}
\includegraphics[width=0.48\textwidth]{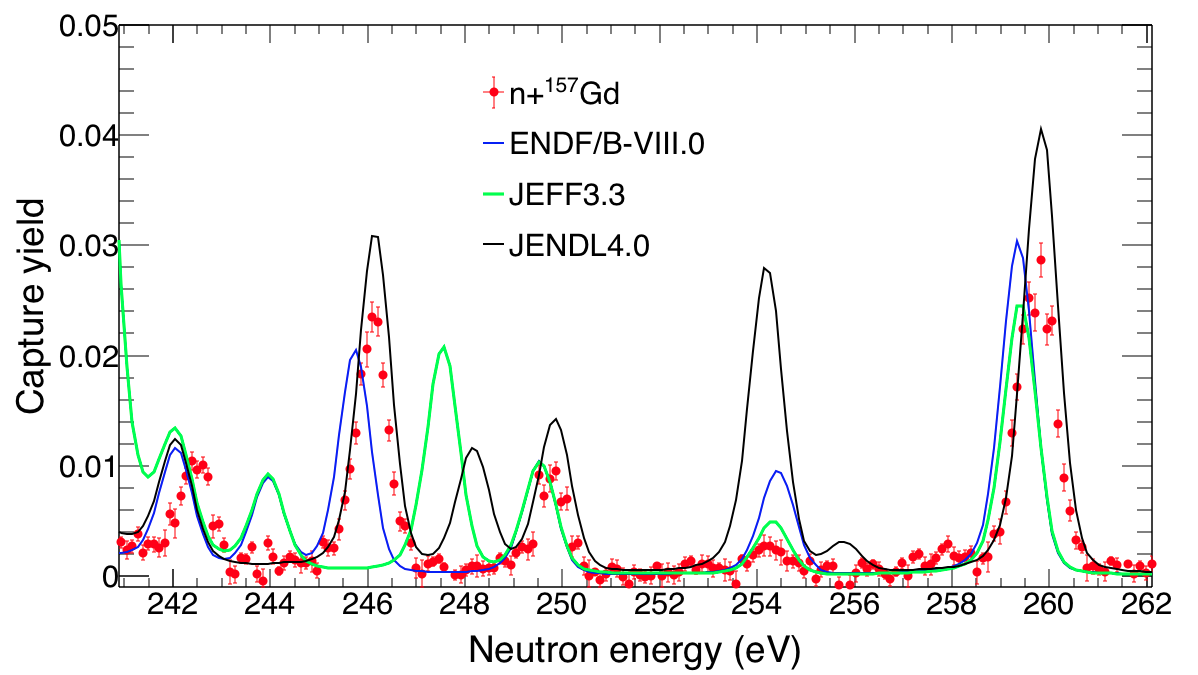}
\caption{(Color online)  $^{157}$Gd(n,$\gamma$) capture yield from the present work and comparison with the expected capture yield, calculated on the basis of the cross sections in ENDF/B-VIII.0, JEFF-3.3 and JENDL-4.0 libraries \label{fig:Gd7}}
\end{figure}

\begin{figure}
\includegraphics[width=0.48\textwidth]{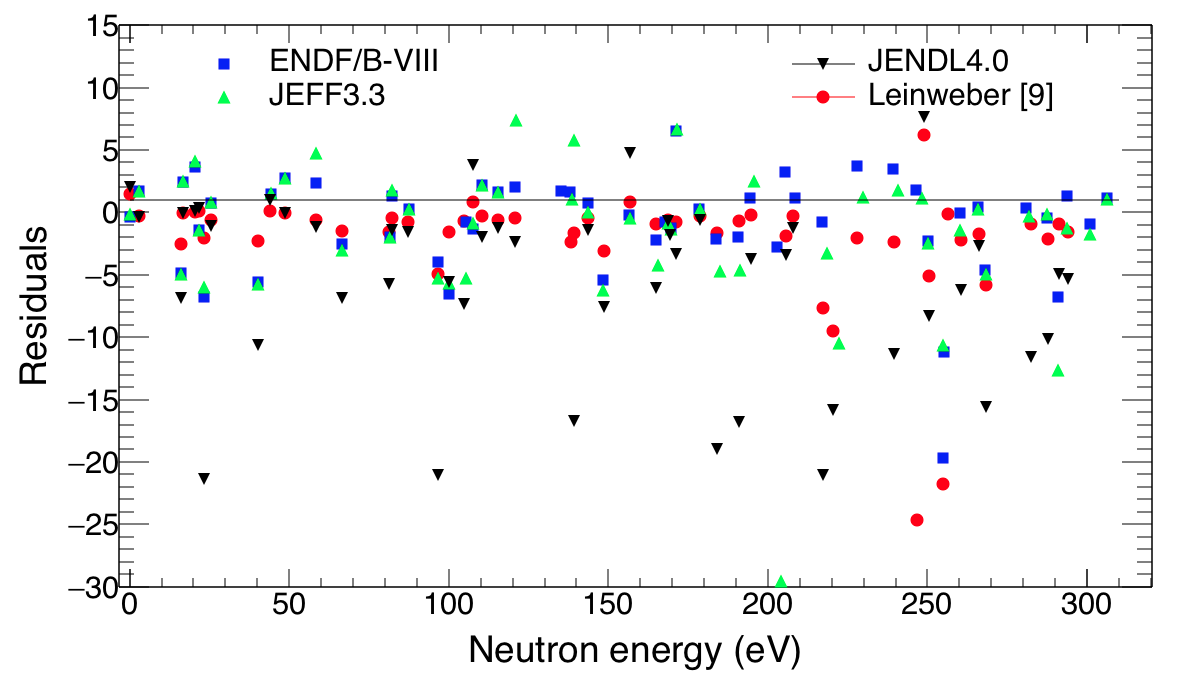}
\caption{(Color online) $^{157}$Gd(n,$\gamma$) residuals between resonance kernel of this work and ENDF/B-VIII.0, JEFF-3.3 and JENDL-4.0 evaluations and  Ref.~\cite{LEINW}, as a function of neutron resonance energy. \label{fig:kernel157}}
\end{figure}

At higher energies, evaluations show differences and inconsistencies. For instance the spin of the first resonance at 0.032 eV is $J=2$ in ENDF/B-VIII.0 and JENDL-4.0 while it is  $J=1$ in JEFF-3.3. Moreover, the average $\Gamma_\gamma$ width is 91, 99 and 117 meV, in these libraries, respectively. There are also doubtful resonances at 135.19, 137.9, 202.69, 208.5, 255.2, 300.9 and 306.4 eV, present in the ENDF/B-VIII.0 evaluation (60 resonances in total) which are neither reported in JEFF-3.3 (which contains 56 resonances ) nor in JENDL-4.0 (with 54 resonances). Figure \ref{fig:Gd7} shows the energy regions where largest discrepancies are present. The present data confirm the resonances in the ENDF/B-VIII.0 evaluation, with the exception of the resonances at 139 eV and 206 eV, moreover the resonance at 220.65 in  ENDF/B-VIII.0 is rather a doublet. 
The results of the resonance shape analysis are summarized in Table~\ref{tab:res157}. Also the correlation coefficient between partial widths $\rho(\Gamma_\gamma, \Gamma_n)$, resulting from the {\sc SAMMY} fit, is reported.
\begin{table}
\caption{Resonances in $^{157}$Gd(n,$\gamma$). Uncertainties are from the fit.\label{tab:res157}}
\begin{ruledtabular}
\begin{tabular}{lcccc}
Energy&$\Gamma_\gamma$ & g$\Gamma_n$ & $\rho$($\Gamma_\gamma$,$\Gamma_n$)&capture kernel\\
(eV)  & (meV)& (meV) & &(meV)\\ 
\hline
0.0314	 &	111.80(2) &	0.2921(1)	&	$-0.11$	&	0.2908(1)	\\
2.8330(1)	&	109.7(3) &	0.2319(4)	&	0.17	&	0.2311(4)	\\
16.218(3)	&	116(7)	 &	0.113(3)	&	0.68	&	0.113(3)	\\
16.7946(2)	&	104.2(5) &	8.74(3)	        &	$-0.49	$&	7.70(1)	\\
20.5262(3)	&	100.3(6) &	8.47(2)	        &	$-0.49$	&	7.47(2)	\\
21.602(2)	&	91(5)	&	0.217(4)	&	0.59	&	0.216(4)	\\
23.290(2)	&	100(4)	&	0.223(3)	&	0.54	&	0.221(3)	\\
25.3653(8)	&	103(2)	&	1.187(7)	&	0.31	&	1.165(7)	\\
40.091(3)	&	101(6)	&	0.368(6)	&	0.44	&	0.365(6)	\\
44.1374(8)	&	106(2)	&	5.98(3)	&	$-0.11$	&	5.48(2)	\\
48.7077(6)	&	103(1)	&	17.2(1)	&	$-0.74$	&	13.56(4)	\\
58.2928(7)	&	109(2)	&	19.9(1)	&	$-0.77$	&	15.44(5)	\\
66.536(1)	&	108(3)	&	4.46(3)	&	$-0.17$	&	4.02(3)	\\
81.312(2)	&	129(4)	&	6.54(5)	&	0.01	&	5.76(5)	\\
82.103(3)	&	102(6)	&	4.10(5)	&	0.40	&	3.86(5)	\\
87.175(2)	&	93(4)	&	6.61(6)	&	$-0.28$	&	6.61(5)	\\
96.572(2)	&	96(4)	&	6.21(7)	&	$-0.51$	&	6.21(4)	\\
100.160(2)	&	130(4)	&	10.7(1)	&	$-0.53$	&	10.7(6)	\\
104.909(2)	&	140(5)	&	14.4(9)	&	$-0.73$	&	14.4(7)	\\
107.370(4)	&	109(8)	&	3.20(5)	&	0.30	&	3.06(5)	\\
110.550(1)	&	99(4)	&	30.6(7)	&	$-0.94$	&	20.5(1)	\\
115.373(2)	&	97(4)	&	13.8(2)	&	$-0.65	$&	11.24(9)	\\
120.861(2)	&	100(1)	&	93(2)	&	$-0.84$	&	37.5(2)	\\
135.36(3)	&	110(10)	&	0.66(4)	&	0.15	&	0.66(4)	\\
138.088(2)	&	122(6)	&	29.1(5)	&	$-0.84$	&	21.0(2)	\\
138.974(6)	&	127(11)	&	5.6(1)	&	0.39	&	5.3(1)	\\
143.736(2)	&	114(6)	&	35.0(9)	&	$-0.93$	&	23.5(2)	\\
148.422(4)	&	114(8)	&	6.15(9)	&	$-0.30	$&	5.38(7)	\\
156.592(3)	&	115(7)	&	11.7(2)	&	$-0.42	$&	10.0(1)	\\
164.910(3)	&	94(7)	&	11.0(3)	&	$-0.80$	&	8.37(9)	\\
168.13(2)	&	146(14)	&	1.16(6)	&	0.21	&	1.15(5)	\\
169.45(1)	&	191(18)	&	1.90(7)	&	0.36	&	1.87(7)	\\
171.408(3)	&	205(7)	&	21.9(3)	&	$-0.60$	&	17.1(1)	\\
178.727(4)	&	104(8)	&	10.7(2)	&	$-0.45$	&	9.2(1)	\\
183.985(4)	&	110(8)	&	8.8(2)	&	$-0.60$	&	7.2(1)	\\
190.789(5)	&	144(10)	&	8.2(1)	&	$-0.29$	&	7.1(1)	\\
194.614(3)	&	125(7)	&	26.3(5)	&	$-0.80$	&	19.7(2)	\\
203.06(1)	&	158(15)	&	2.82(8)	&	0.24	&	2.69(7)	\\
205.63(4)	&	82(8)	&	0.66(5)	&	0.04	&	0.65(5)	\\
207.725(3)	&	88(1)	&	108(3)	&	$-0.75$	&	36.4(3)	\\
217.22(1)	&	162(16)	&	2.66(8)	&	0.22	&	2.55(7)	\\
220.39(4)	&	120(12)	&	0.67(5)	&	0.07	&	0.66(5)	\\
221.38(3)	&	199(20)	&	1.26(7)	&	0.21	&	1.24(7)	\\
228.406(9)	&	163(15)	&	5.3(1)	&	0.11	&	5.1(1)	\\
239.572(4)	&	127(3)	&	82(3)	&	$-0.81$	&	30.1(2)	\\
243.8(1)	&	135(13)	&	0.23(2)	&	0.02	&	0.22(2)	\\
246.75(1)	&	149(14)	&	6.3(2)	&	0.06	&	5.9(1)	\\
250.42(2)	&	147(15)	&	1.73(8)	&	0.11	&	1.68(8)	\\
254.6(1)	&	121(12)	&	0.34(4)	&	0.01	&	0.34(3)	\\
255.10(8)	&	113(11)	&	0.53(5)	&	0.01	&	0.53(5)	\\
260.44(1)	&	222(19)	&	7.3(2)	&	0.07	&	6.7(2)	\\
265.99(2)	&	171(17)	&	3.9(1)	&	0.15	&	3.8(1)	\\
268.43(2)	&	119(12)	&	4.6(2)	&	0.04	&	4.3(2)	\\
282.015(6)	&	128(11)	&	20.0(7)	&	$-0.86$	&	14.1(2)	\\
287.73(1)	&	203(19)	&	8.2(2)	&	0.20	&	7.7(2)	\\
291.072(9)	&	147(13)	&	11.5(3)	&	$-0.42$	&	9.5(2)	\\
294.060(6)	&	106(9)	&	24(1)	&	$-0.90$	&	15.1(2)	\\
301.353(8)	&	184(15)	&	13.8(3)	&	$-0.42$	&	11.5(2)	\\
306.92(4)	&	226(22)	&	2.0(1)	&	0.13	&	2.0(1)	\\
\end{tabular}
 \end{ruledtabular}
 \end{table}
A comparison of the kernels from the present analysis to the ones from evaluations and Ref.~\cite{LEINW} is reported as a function of resonance energy in Fig.~\ref{fig:kernel157} in terms of residuals.
On average a good agreement was found with ENDF/B-VIII.0 and JEFF-3.3 evaluations, since the statistical distribution of the ratios was Gaussian with mean 0.98. On the contrary, the comparison with JENDL-4.0 and the data from Leinweber and collaborators does not tend to a Gaussian distribution and the average deviation is 13\% and 18\%, respectively.  From the figure it results that the resonance parameters form JENDL-4.0 are not directly taken from Ref.~\cite{LEINW}.
 
The resonances and structures observed in the energy region above the upper limit of evaluations are reported in Tab.~\ref{tab:new57} in Appendix~\ref{App:newres} together with their capture kernel. 

\subsection{Statistical properties of  neutron resonances\label{SeC_stat}}
Resolved resonance parameters from this analysis, see Tabs. \ref{tab:res155} and \ref{tab:res157} can be used to determine basic statistical properties of resonances. Since we do not see any significant difference in the number of observed resonances with respect to other experiments reported in literature, the estimation of quantities describing the statistical properties of neutron resonances -- $s$-wave neutron strength function, $S_0$, resonance radiative width, $\Gamma_\gamma$, and $s$-wave average resonance spacing, $D_0$ -- should not differ significantly from nuclear data libraries. 

\subsubsection{Neutron strength function}

An estimate of the $s$-wave neutron strength function $S_0$ can be made from the reduced neutron widths as
\begin{equation}
S_0 = \frac{1}{\Delta E} \sum\limits_{\Delta E} g_J \Gamma_n^0
\end{equation}
where $\Delta E$ is the interval of neutron energies which the reduced neutron widths $\Gamma_n^0$ are summed over. The sum goes over resonances of both spins.

Assuming the neutron strength function for $p$-wave resonances is close to the literature values $S_1 = 3.7(1.1)\times 10^{-4}$ for $^{155}$Gd and $S_1 = 2.2(7)\times 10^{-4}$ for $^{157}$Gd \cite{MUGH}, no $p$-wave resonance should be observable in our data as these resonances are too weak. On the other hand, as already pointed out in Ref.~\cite{BARA} the Porter-Thomas (PT) fluctuations of individual neutron widths almost surely prevent observation of some $s$-wave resonances in Gd isotopes. Nevertheless, the contribution of these unobservable resonances to the sum  is very small,  about 1\% in RRR for both nuclei.   

The uncertainty in $S_0$ is given by the uncertainty in individual $\Gamma_n^0$ values from {\sc SAMMY} fitting and by the expected PT fluctuations which the $\Gamma_n^0$ values are expected to follow. The PT fluctuation adds an uncertainty $\sqrt{2/N_R} S_0$, where $N_R$ is the number of resonances.
Our data yield $S_0 = 2.01(28) \times 10^{-4}$ and $2.17(40) \times 10^{-4}$ for $^{155}$Gd and $^{157}$Gd determined from energy regions below 180 and 300 eV, respectively. These values agree with values available in literature: $1.99(28)\times 10^{-4}$ \cite{BARA} and $2.20(14)\times 10^{-4}$ \cite{MUGH} for $^{155}$Gd and  $2.20(40)\times 10^{-4}$ \cite{MUGH} for $^{157}$Gd.  The dominant contribution to listed uncertainty comes from the Porter-Thomas fluctuations.
 
\begin{figure}
\includegraphics[clip,width=\columnwidth]{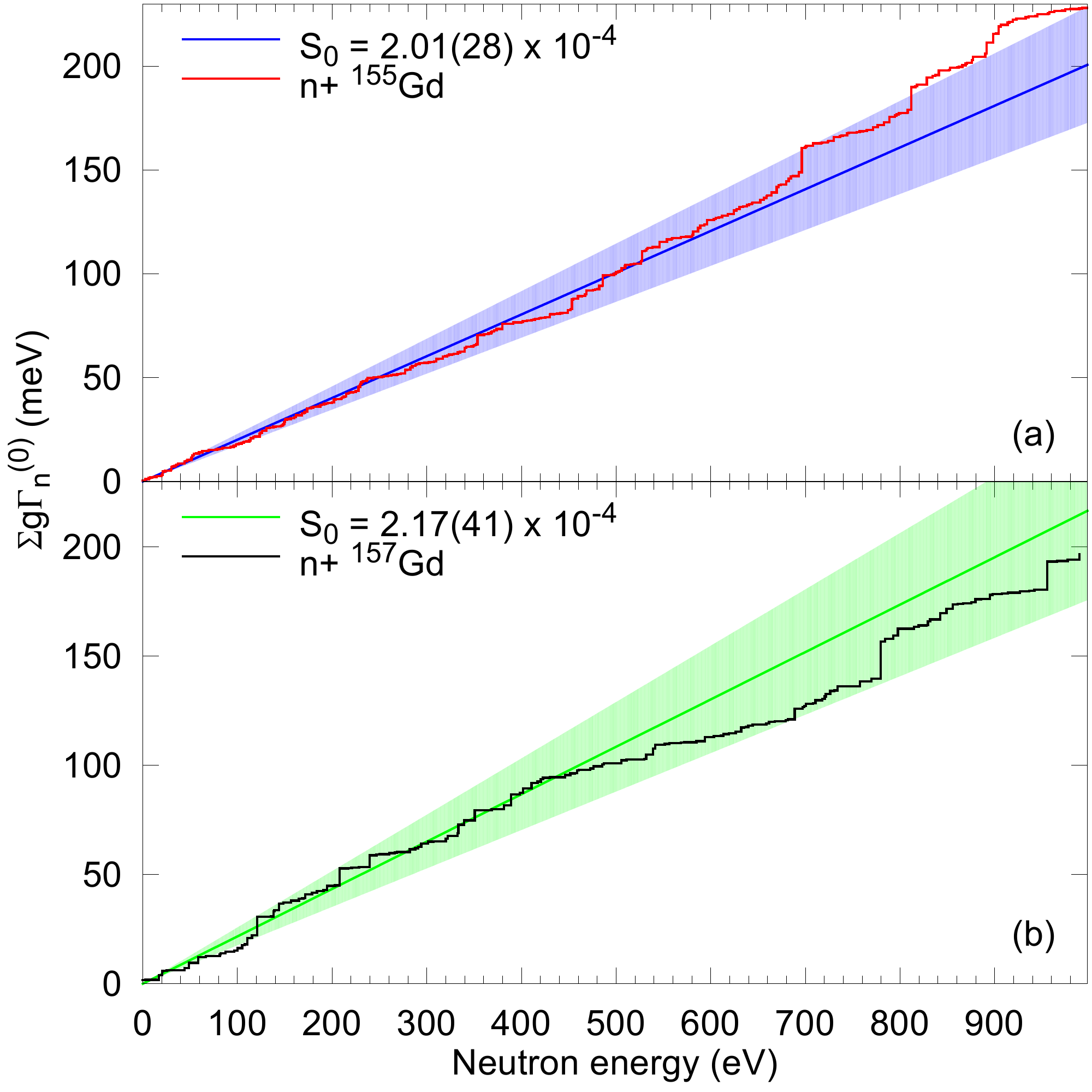}
\caption{\label{fig:s0} Cumulative distribution of reduced neutron widths for both Gd nuclei. Solid lines correspond to $S_0=2.01\times 10^{-4}$  for $^{155}$Gd and $S_0=2.17\times 10^{-4}$  for $^{155}$Gd, respectively. Coloured regions  indicate expected corridor for both isotopes due to Porter-Thomas fluctuations.}
\end{figure}

 Figure \ref{fig:s0} shows the dependence of $\sum g_J \Gamma_n^0$ on neutron energy. For resonances where only resonance kernel is given in Tabs. \ref{tab:new55} and \ref{tab:new57}, we assumed $\Gamma_\gamma=108$ and 105 meV for $^{155}$Gd and $^{157}$Gd, respectively, and spin was assigned randomly assuming that the ratio of number of $J=1$ to $2$ resonances is $3/5$ as expected from standard spin dependence of the level density. Expected uncertainties, corresponding to the average $\pm$ rms from Porter-Thomas distribution are also indicated in the figure; those for $^{157}$Gd are higher due to larger resonance spacing. 

\subsubsection{Total radiation width}

The radiation widths from resonance shape analysis are very precisely determined at low $E_n$ but their uncertainty significantly increases with neutron energy. 
The statistical model predicts that due to the many possible decay channels the $\Gamma_\gamma$ should not vary much for resonances in a given isotope. This quantity is also expected to depend only weakly on the resonance spin. Simulations of the $\gamma$ decay of resonances using the {\sc DICEBOX} code indicated that the fluctuation of $\Gamma_\gamma$ are expected to be for resonances with the same spin about 1-2\% with a realistic model of nuclear level density and of photon strength functions~\cite{Baramsai2013,Chyzh2011}. The simulations also predicted similar difference, about 2\%, between the $\Gamma_\gamma$ expectation values between $J=1$ and 2 resonances.

Assuming Normal distribution of actual $\Gamma_\gamma$ values we tried to estimate the mean value and the width of the distribution of this quantity using the maximum-likelihood (ML) method. Uncertainties of individual values from {\sc SAMMY} fit were taken into account in determining the parameters of this distribution. 
Using resonances for $E_n < 50$ eV (i.~e. the region where the Doppler broadening does not dominate the observed widths of the resonances), the ML method yielded the expectation value ${\overline \Gamma_\gamma} = 109(2)$ and 105(2) meV. 



\subsubsection{Resonance spacing}

\begin{figure}
\includegraphics[clip,width=\columnwidth]{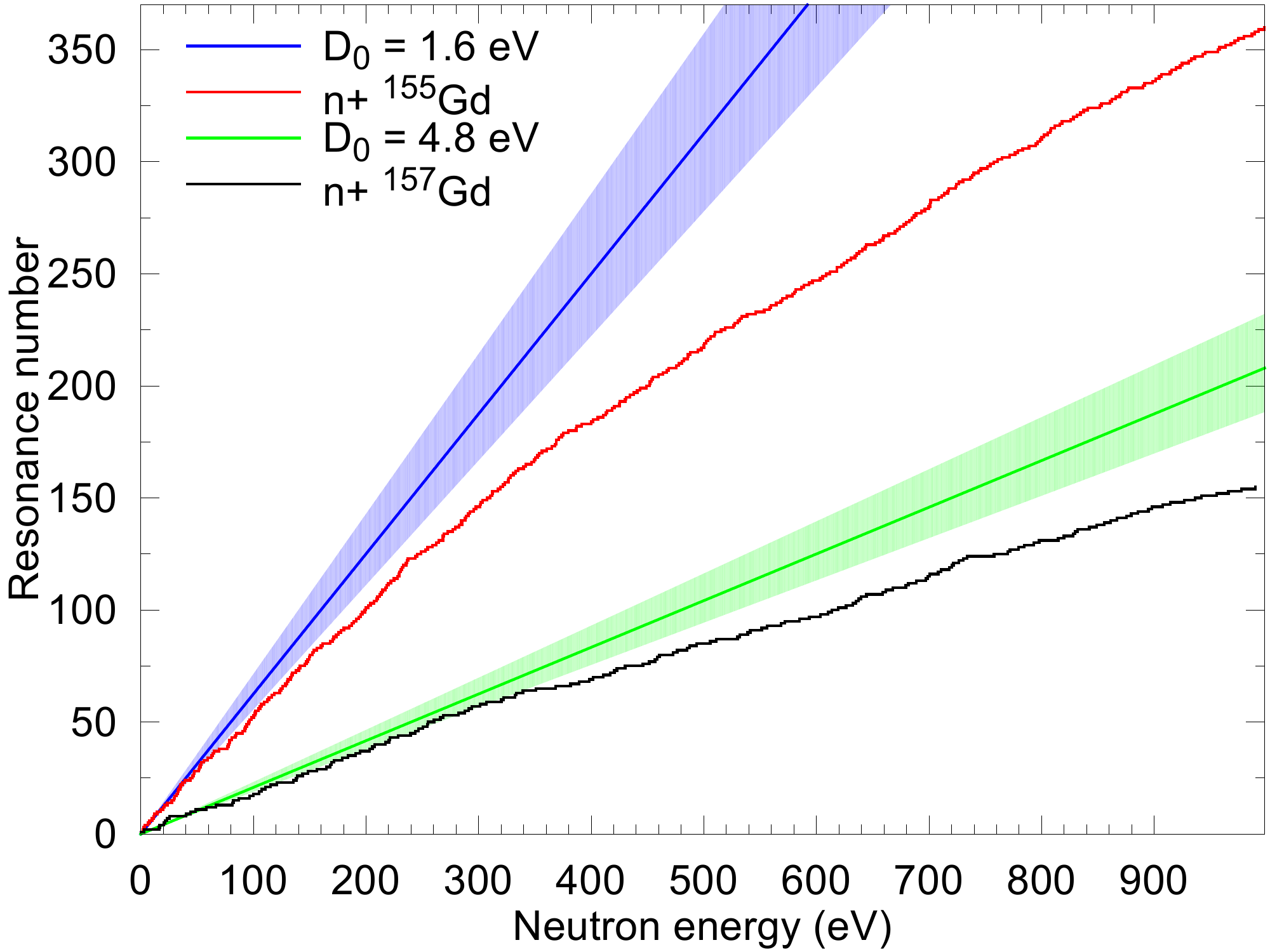}
\caption{\label{fig:spacing} Cumulative distribution of observed resonances for both Gd isotopes. Blue and green lines indicate expected number of levels for the most probable values of spacing. Coloured regions indicate expected corridor for both isotopes.}
\end{figure}

The cumulative plot of the number of resonances as a function of neutron energy is shown in Fig. \ref{fig:spacing}. The observed deviation from a straight line at higher energies clearly indicates an increasing number of missing levels. We should remind the reader that we are not sure if the reported structures above 180 eV and 300 eV in $^{155}$Gd and $^{157}$Gd, respectively, correspond to individual resonances.  
In reality, as mentioned above the PT fluctuations of individual neutron widths almost surely prevent observation of all resonances in Gd isotopes from very low neutron energies. The resonance spacing thus cannot be calculated as a simple ratio $\Delta E / N_{obs}$, where $N_{obs}$ is the number of observed resonances, but must be corrected. Many different ways of correction have been applied in the past, see e.~g. Refs.~\cite{MITCH, BARA} . In this work we tried to estimate the spacing using comparison of the observed number of resonances above an assumed threshold applied to the resonance kernel with predictions of statistical model calculations. 

Several thousands of artificial resonance sequences were generated using the above-given values of $S_0$ and $\Gamma_\gamma$. The number of observed resonances is for several different thresholds and maximum neutron energies (below 400 eV) nicely consistent with $D_0$ in the range about $1.4-1.8$ eV and about $4.3-5.3$ eV in $^{155}$Gd and $^{157}$Gd, respectively. These values are fully consistent with values available in the literature. Use of higher neutron energies for comparison starts to be problematic as reported structures may correspond to close resonance multiplets.

\section{Summary and conclusion \label{Sec_6}}
The measurement of the yield of the $^{155}$Gd(n,$\gamma$) and $^{155}$Gd(n,$\gamma$) reaction for $E_n<1$ keV is described. These data sets, which will be submitted to the EXFOR database, can be used for future evaluations, hopefully in combination with the results of a new transmission experiment.  From the R-matrix analysis of the present data, we extracted resonance parameters and therefore cross sections from thermal energy to about 1 keV.

 The comparisons with ENDF/B-VIII.0 and JEFF-3.3 nuclear data libraries show a fair agreement in the resonance region, whereas sizable differences are found with respect to the experiment by Leinweber and collaborators and therefore with the JENDL-4.0 evaluation. 
 
 The thermal cross sections extracted in this work are about 2\% higher for $^{155}$Gd and  6\% smaller for $^{157}$Gd than those reported in nuclear data libraries. For $^{155}$Gd the present data is consistent with the data of Ohno and Leinweber and is compatible with the data of M{\o}eller within 1.2 standard deviations. For $^{157}$Gd the present data is consistent with the data of Ohno is compatible with the data of M{\o}eller and Leinweber within 1.2 and 1.5 standard deviations, respectively.
\begin{acknowledgments}
The isotopes used in this research were supplied by the United States Department of Energy Office of Science by the Isotope Program in the Office of Nuclear Physics. 

This research was partially funded by the European Atomic Energy Community (Euratom) Seventh Framework Programme FP7/2007-2011 under the Project CHANDA (Grant No. 605203).

We acknowledge support from FPA2014-52823-C2-1-P (MINECO)
\end{acknowledgments}
\clearpage
\newpage

\appendix
\section{$^{155}$Gd and $^{157}$Gd resonances not included in evaluations\label{App:newres}}
\begin{figure}
\includegraphics[width=0.48\textwidth]{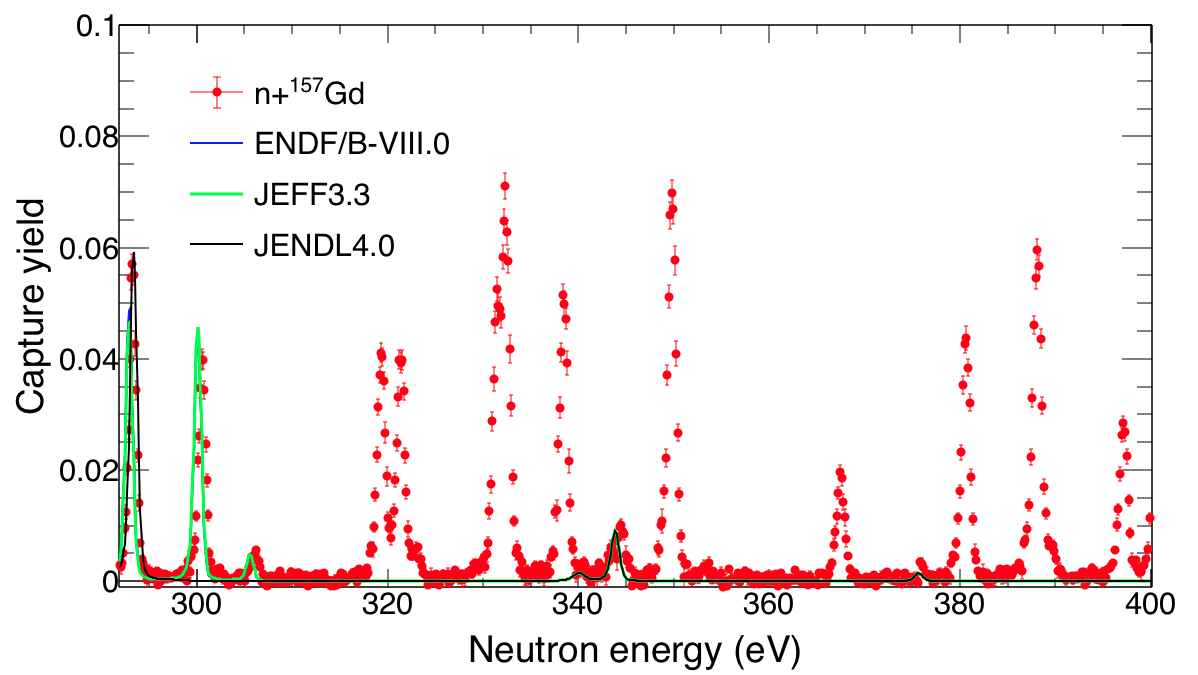}
\includegraphics[width=0.48\textwidth]{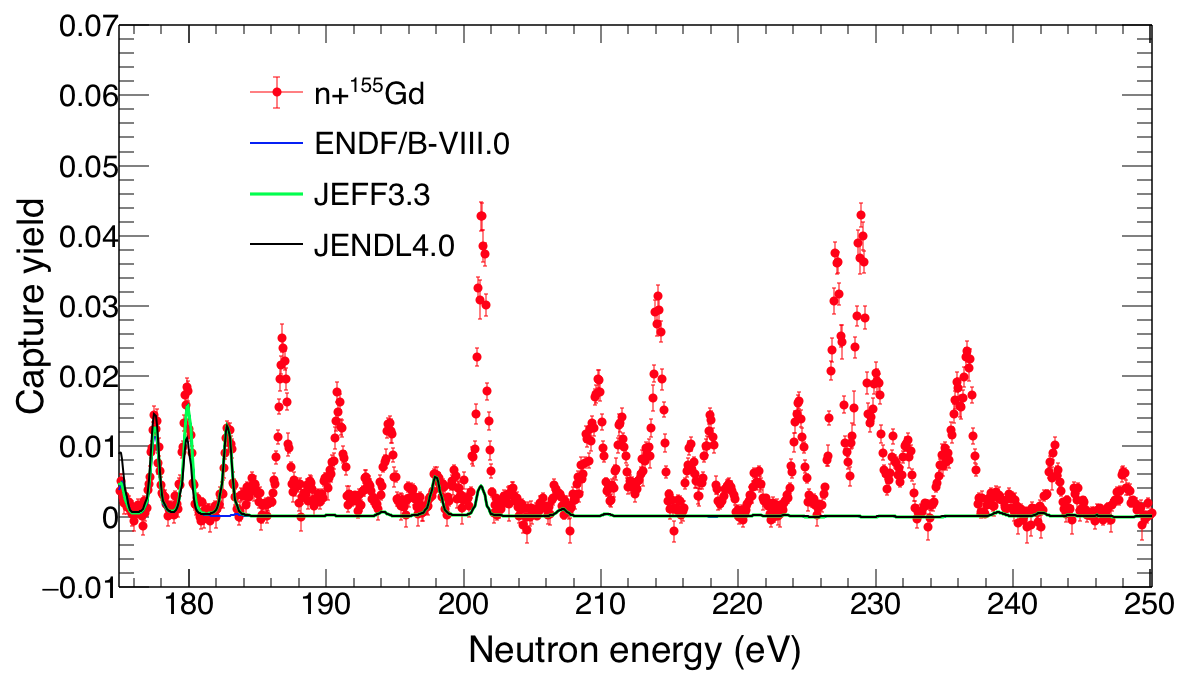}
\caption{(Color online)  $^{155,157}$Gd(n,$\gamma$) capture yield from the present work and calculated according to ENDF/B-VIII.0, JEFF-3.3 and JENDL-4.0 libraries around the boundary of RRR.\label{fig:GdURR}}
\end{figure}

The present data clearly show structures in the $^{155}$Gd(n,$\gamma$) and $^{157}$Gd(n,$\gamma$) cross sections also above the resolved resonance region reported in the evaluations, as can be seen in Fig.~\ref{fig:GdURR}.

\begin{figure}
\includegraphics[width=0.48\textwidth]{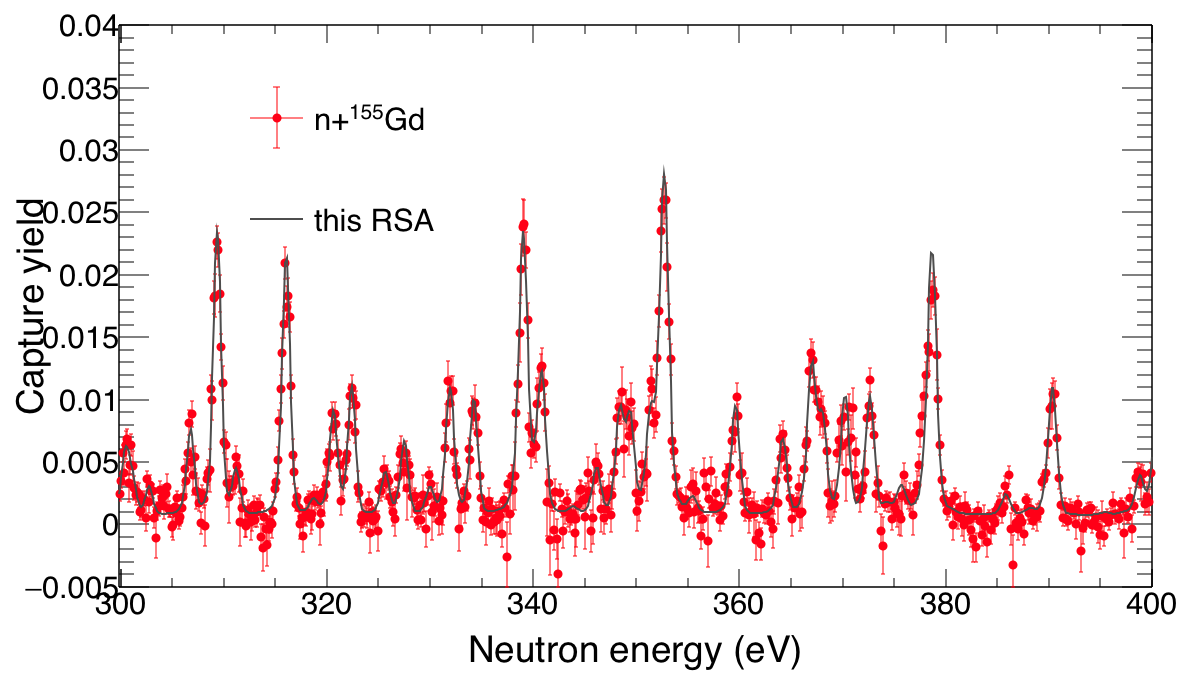}
\includegraphics[width=0.48\textwidth]{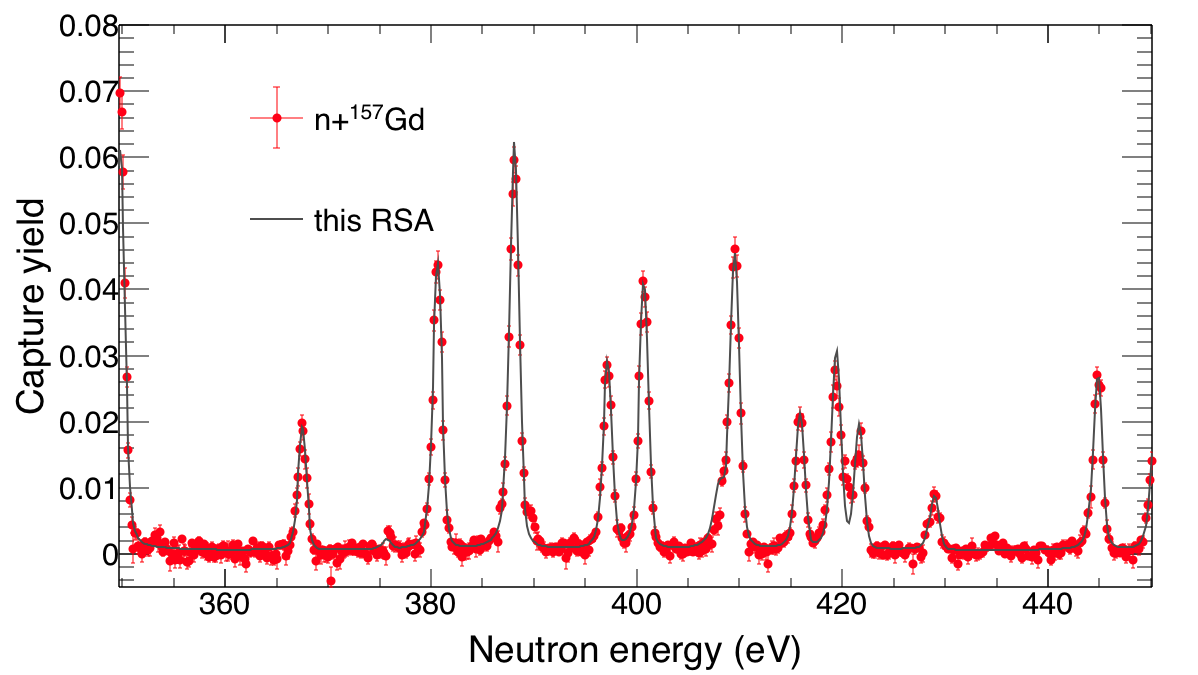}
\caption{(Color online)  $^{155,157}$Gd(n,$\gamma$) capture yield from the present experiment and the results of a  resonance shape analysis.\label{fig:nuove}}
\end{figure}

The properties of these structures (267 for $^{155}$Gd and 96 for $^{157}$Gd), namely resonance energy and area, are reported in Table~\ref{tab:new55} and~\ref{tab:new57}  up to $E_n=1$ keV. The resonance analysis was performed with {\sc SAMMY}, by adopting a constant capture width ${\overline \Gamma_\gamma} = 109(2)$ and 105(2) meV for  $n+^{155}$Gd(n,$\gamma$) and  $n+^{157}$Gd(n,$\gamma$) reaction, respectively, and assuming angular orbital momentum $\ell=0$. Examples of the quality of the resonance shape analysis are showed in Fig.\ref{fig:nuove}.
 \begin{longtable*}{@{\extracolsep{\fill}}lc | lc | lc | lc@{}}
\caption{Some properties of the 272 $^{155}$Gd(n,$\gamma$) resonances not included in the evaluations. Uncertainties are from the fit.\label{tab:new55}}\\
\hline 
\hline
 Energy & capture kernel &Energy & capture kernel  &Energy & capture kernel &Energy & capture kernel \\
(eV) & (meV) &(eV) & (meV) &(eV) & (meV) &(eV) & (meV)\\
\hline
\endfirsthead
\multicolumn{8}{c}{ \tablename\ \thetable{}  (Continued) } \\ \hline
 Energy & capture kernel &Energy & capture kernel  &Energy & capture kernel &Energy & capture kernel \\
(eV) & (meV) &(eV) & (meV) &(eV) & (meV) &(eV) & (meV)\\
\hline
\endhead
\hline \multicolumn{8}{c}{{Continued on next page}} \\ 
\endfoot
\hline \hline
\endlastfoot
185.08(4) & 1.33(9)& 328.37(4) & 4.3(3)& 526.20(4) & 20.7(1)& 738.4(2) & 6.0(5)\\
187.36(1) & 8.22(2)& 330.86(12)& 1.5(1)& 527.37(3) & 30.2(8)& 739.8(3) & 1.5(1)\\
189.05(5) & 1.08(8)& 332.83(3) & 7.5(4)& 530.35(9) & 4.9(4)& 744.05(6) & 21(1)\\
191.37(1) & 5.6(2)& 335.14(3) & 7.0(4)& 532.14(5) & 13.4(7)& 745.9(2) & 5.2(4)\\
193.45(3) & 1.45(9)& 339.94(2) & 18.7(6)& 533.44(6) & 10.1(6)& 751.7(2) & 3.9(3)\\
195.10(1) & 4.3(1)& 341.75(4) & 8.3(4)& 538.41(6) & 7.3(5)& 754.4(3) & 2.8(3)\\
196.68(5) & 1.01(7)& 347.08(9) & 2.9(2)& 545.97(3) & 32(1)& 757.4(1) & 9.6(7)\\
199.1(2) & 0.12(1)& 349.34(6) & 6.3(4)& 553.68(4) & 16.3(7)& 760.9(2) & 5.4(4)\\
199.89(4) & 1.49(9)& 350.32(6) & 6.0(4)& 558.5(3) & 7.5(6)& 764.92(8) & 12.2(8)\\
201.87(1) & 14.2(3)& 352.30(6) & 5.6(4)& 559.6(1) & 4.9(4)& 771.73(7) & 19.3(9)\\
203.97(5) & 0.98(8)& 353.65(2) & 27.7(6)& 564.4(2) & 2.2(2)& 776.49(6) & 20(1)\\
207.20(5) & 0.83(7)& 356.40(0) & 1.0(1)& 568.28(6) & 9.3(6)& 778.9(3) & 2.6(2)\\
209.47(3) & 3.0(2)& 360.58(3) & 7.5(4)& 569.9(3) & 0.72(7)& 783.16(5) & 23(1)\\
210.33(1) & 7.2(2)& 365.16(5) & 5.2(3)& 573.7(1) & 2.7(2)& 788.80(5) & 26.9(8)\\
212.04(2) & 4.8(2)& 367.99(3) & 10.2(5)& 578.8(1) & 4.3(3)& 794.63(9) & 12.9(9)\\
213.64(3) & 2.6(2)& 368.9(3) & 0.40(4)& 580.48(7) & 12.4(8)& 796.5(1) & 12.1(8)\\
214.71(1) & 12.5(3)& 369.03(6) & 6.5(4)& 581.40(5) & 19.4(9)& 798.8(1) & 11.2(8)\\
217.10(2) & 2.9(1)& 371.23(4) & 8.5(4)& 586.61(4) & 21.2(8)& 800.2(3) & 2.9(3)\\
218.48(2) & 5.2(2)& 373.66(3) & 8.7(4)& 588.92(4) & 17.1(8)& 802.1(4) & 1.7(2)\\
219.88(4) & 1.4(1)& 375.3(2) & 0.67(6)& 593.29(6) & 9.6(5)& 805.4(4) & 0.47(5)\\
221.99(3) & 2.1(1)& 379.68(2) & 22.2(6)& 596.38(4) & 23.0(8)& 807.8(3) & 5.7(5)\\
224.99(1) & 6.7(2)& 386.9(2) & 0.87(8)& 603.40(6) & 9.9(7)& 808.20(9) & 24(1)\\
227.77(1) & 17.3(3)& 386.9(2) & 0.68(7)& 605.53(6) & 10.8(7)& 811.81(7) & 36.7(4)\\
229.46(2) & 12.0(6)& 391.40(3) & 10.2(5)& 608.7(3) & 7.0(8)& 816.3(2) & 9.5(7)\\
229.63(4) & 7.0(5)& 399.89(8) & 3.0(2)& 612.0(3) & 13(1)& 818.0(1) & 13.4(8) \\
230.68(2) & 8.2(3)& 401.56(3) & 8.2(4)& 616.85(7) & 9.5(7)& 825.9(4) & 0.54(5) \\
231.90(3) & 2.9(2)& 405.97(5) & 5.0(3)& 618.8(2) & 1.5(1)& 828.28(6) & 29.5(8)\\
232.91(2) & 4.2(2)& 410.67(6) & 5.2(3)& 622.1(3) & 1.1(1)& 831.4(4) & 1.2(1)\\
235.52(3) & 3.1(2)& 412.96(7) & 3.4(3)& 624.50(4) & 20.5(8)& 834.44(9) & 14.2(8)\\
236.48(2) & 6.9(3)& 414.22(4) & 8.4(5)& 627.22(6) & 12.8(7)& 837.3(2) & 8.4(7)\\
237.35(1) & 10.4(3)& 418.50(9) & 2.7(2)& 630.55(4) & 16.5(6)& 841.01(7) & 25.6(9)\\
243.63(2) & 3.9(2)& 420.53(8) & 4.7(3)& 634.2(1) & 4.8(4)& 851.7(2) & 7.3(6)\\
245.35(7) & 0.98(8)& 425.1(1) & 4.4(3)& 635.6(4) & 0.92(9)& 853.07(9) & 16.3(9)\\
248.74(3) & 2.5(1)& 425.2(3) & 1.7(2)& 638.10(6) & 14.0(8)& 860.60(3) & 4.3(4)\\
252.8(1) & 0.77(7)& 430.07(2) & 19.1(7)& 640.69(9) & 8.0(5)& 862.3(2) & 7.7(6)\\
254.79(2) & 5.9(2)& 430.8(1) & 3.0(3)& 643.3(1) & 8.1(6)& 865.92(8) & 16.8(9)\\
259.15(3) & 2.4(2)& 434.4(2) & 1.2(1)& 644.13(8) & 9.6(6)& 869.3(1) & 15.3(9)\\
262.51(3) & 2.9(2)& 437.61(7) & 4.4(3)& 652.36(4) & 21.9(9)& 870.8(2) & 9.3(7)\\
264.84(3) & 3.5(2)& 440.98(7) & 3.8(3)& 656.26(6) & 14.8(7)& 875.96(8) & 25(1)\\
268.4(2) & 1.0(1)& 443.4(1) & 2.4(2)& 659.17(7) & 17(1)& 876.97(9) & 21(1)\\
268.38(9) & 2.9(2)& 449.21(2) & 21.4(7)& 659.6(2) & 7.1(6)& 890.1(1) & 23(1)\\
269.37(4) & 2.6(2)& 452.03(8) & 4.8(3)& 664.19(4) & 23.7(9)& 891.45(6) & 48(1)\\
272.34(4) & 2.7(2)& 452.1(4) & 4.8(4)& 669.76(5) & 25.9(8)& 898.44(7) & 31.5(7)\\
276.96(1) & 17.4(4)& 453.5(5) & 29.1(9)& 671.66(6) & 17(1)& 901.2(1) & 1.9(2)\\
279.27(6) & 1.8(1)& 454.7(3) & 6.4(5)& 674.0(1) & 8.1(6)& 904.29(6) & 29.8(8)\\
282.52(1) & 13.5(3)& 459.75(3) & 19.2(7)& 677.2(4) & 0.86(8)& 906.5(4) & 15(1)\\
285.17(3) & 6.0(3)& 463.8(2) & 1.8(2)& 679.86(5) & 19.6(8)& 913.91(9) & 22(1)\\
284.34(3) & 6.5(3)& 467.19(4) & 12.5(6)& 682.45(4) & 24(1)& 915.43(4) & 8.6(7)\\
288.09(2) & 7.3(3)& 468.62(3) & 24.2(8)& 684.84(5) & 17.0(7)& 919.4(1) & 23(1)\\
288.99(4) & 4.8(3)& 475.70(8) & 5.2(4)& 686.8(3) & 1.4(1)& 923.0(2) & 11.6(8)\\
290.85(7) & 1.2(1)& 477.97(6) & 6.5(5)& 688.4(4) & 0.94(9)& 924.7(5) & 0.60(6)\\
292.36(2) & 5.9(3)& 480.58(9) & 3.8(3)& 693.22(4) & 29(1)& 930.4(5) & 0.46(5)\\
295.71(8) & 1.3(1)& 482.26(2) & 22.6(8)& 696.36(6) & 36.7(4)& 932.9(1) & 12.6(8)\\
297.7(1) & 0.99(9)& 485.6(2) & 1.7(2)& 699.74(5) & 8.3(7)& 935.9(1) & 13.3(9)\\
301.25(5) & 3.0(2)& 487.97(3) & 30.1(6)& 700.7(3) & 4.9(4)& 942.8(4) & 2.5(2)\\
301.93(8) & 1.8(2)& 485.9(3) & 1.3(1)& 701.0(2) & 8.9(7)& 945.3(1) & 20(1)\\
303.57(7) & 1.5(2)& 494.8(1) & 2.9(2)& 701.4(4) & 2.4(2)& 956.5(2) & 7.2(6)\\
307.60(4) & 4.5(3)& 497.77(5) & 10.0(5)& 708.28(5) & 21(1)& 958.0(2) & 16(1)\\
310.20(1) & 16.0(4)& 499.99(9) & 6.9(6)& 711.3(4) & 1.3(1)& 963.2(1) & 13.4(8)\\
311.98(8) & 2.2(2)& 500.17(8) & 7.7(7)& 713.4(4) & 0.56(6)& 968.1(5) & 0.22(2)\\
312.3(3) & 0.36(4)& 502.18(7) & 6.0(4)& 716.6(2) & 3.7(3)& 971.5(4) & 1.3(1)\\
316.88(1) & 15.0(4)& 503.77(6) & 6.6(4)& 717.5(2) & 4.6(4)& 974.6(1) & 15(1)\\
319.5(2) & 0.72(7)& 505.86(3) & 17.5(6)& 723.5(1) & 9.0(6)& 979.6(2) & 5.6(5)\\
321.50(3) & 5.9(3)& 509.43(4) & 16.0(9)& 724.1(2) & 7.1(6)& 986.7(5) & 0.10(1)\\
323.27(2) & 7.4(3)& 510.02(6) & 11.3(7)& 726.6(4) & 0.90(9)& 989.4(2) & 6.0(5)\\
326.4(2) & 1.4(1)& 515.3(1) & 4.1(3)& 730.15(4) & 30(1)& 992.6(2) & 8.0(7)\\
326.6(2) & 1.2(1)& 518.82(7) & 5.5(4)& 736.6(1) & 8.4(6)& 997.58(7) & 2.4(2)\\
 \\hline
 \end{longtable*}
\begin{table*}
\caption{Some properties of the 96 $^{157}$Gd(n,$\gamma$) resonances not included in the evaluations. Uncertainties are from the fit.\label{tab:new57}}
\begin{ruledtabular}
\begin{tabular}{lc | lc | lc | lc}
 Energy & capture kernel &Energy & capture kernel  &Energy & capture kernel &Energy & capture kernel \\
(eV) & (meV) &(eV) & (meV) &(eV) & (meV) &(eV) & (meV)\\
\hline
320.19(1)& 17.6(3) & 487.28(3)& 6.6(3)  & 658.71(4)	        &  11.5(5) &   814.73(4)     & 17.6(8) \\
322.23(1)& 14.034(2)& 493.59(3)& 0.87(08) & 661.62(4)	& 14.9(6) &  819.84(5)	& 14.4(7) \\
332.30(2)& 17.55(6) & 505.67(1)& 17.9(4)  & 667.67(6)	& 5.6(4) &  828.1(2)     & 3.8(3) \\
333.29(1)& 25.2(5)    & 511.02(4)& 6.7(3)  & 679.42(5)	& 11.4(6) &  829.62(4)	& 27.9(9) \\
339.37(1)& 23.9(4)  & 529.90(3)& 6.4(3) & 681.69(6)	         & 8.8(5)  &  831.84(5)	& 19.0(8) \\
350.74(1)& 27.0(4)     & 531.87(2)& 26.8(5)  & 688.84(2)	& 42.8(9)    &  842.40(3)	& 37.0(9) \\
368.48(2)& 9.3(3) & 538.99(2)& 24.6(5)    & 697.15(4)	& 14.1(5)  &  849.02(4)	& 31(1) \\
381.67(1)& 18.7(3)  & 541.33(2)& 25.1(6)  & 698.73(6)	& 12.3(6) &  855.72(3)	& 30.8(9) \\
389.16(1)& 39.5(5)    & 551.33(3)& 8.6(3)  & 700.64(4)	& 17.8(7)  &  861.1(1)     & 8.2(5) \\
398.24(1)& 12.1(3) & 556.16(4)& 7.8(3) & 708.24(7)	        & 5.5(3) &  866.7(1)     & 5.1(4) \\
401.78(1)& 24.1(4)  & 568.49(8)& 3.0(2)& 710.64(2)	        & 19.5(6)  &  875.46(8)	& 12.8(7) \\
410.67(1)& 22.2(4)  & 572.13(5)& 5.1(3) & 718.04(5)	        & 11.8(6)  &  879.67(3)	& 24.7(8) \\
416.97(2)& 12.2(4) & 584.93(4)& 8.7(4) & 720.49(5)	        & 21.5(9)  &  885.6(1)     & 6.3(5) \\
420.57(1)& 14.1(3)  & 593.79(2)& 21.3(5)  & 721.59(4)	& 20.8(9)  &  894.79(4)	& 30(1) \\
422.78(2)& 11.4(3) & 603.41(3)& 12.1(5) & 726.31(3)	& 23.9(7)  &  897.79(7)	& 11.0(6) \\
430.07(3)& 5.0(2) & 610.44(9)& 2.4(2) & 730.52(8)	& 5.2(4) &  907.48(4)	& 5.5(5) \\
446.05(1)& 13.7(3) & 613.54(3)& 12.4(4) & 733.88(3)	& 29.0(8)  &  914.08(9)	& 7.6(4) \\
451.64(2)& 11.8(4) & 619.32(5)& 6.5(4) & 757.80(2)	& 31.0(8)  &  926.70(7)	& 14.2(7) \\
456.75(4)& 4.4(2)& 626.50(3)& 14.9(5) & 769.81(3)	& 23.5(7)  &  936.46(1) 	& 7.5(5) \\
458.72(1)& 16.0(3)  & 632.22(2)& 21.8(5)  & 771.0(2) 	& 3.2(3) &  941.99(5)	& 12.9(6) \\
460.78(1)& 2.6(3)& 635.00(8)& 3.8(3) & 779.48(3)	& 56.2(5)   &  955.65(3)	& 55.2(7) \\
472.70(2)& 11.6(4) & 637.29(1)& 2.6(2)& 784.34(3)	& 22.6(8)  &  965.91(1) 	& 6.9(5) \\
476.40(1)& 14.5(3)  & 639.72(3)& 11.5(4) & 792.90(3)	& 25.8(8)  &  977.24(8)	& 13.9(8) \\
485.29(2)& 15.8(4) & 644.47(4)& 8.8(4)  & 797.78(3)	& 37.9(9)    &  989.79(4)	& 35(1) \\
\end{tabular}
 \end{ruledtabular}
 \end{table*}
%
%
%
%

\end{document}